\documentclass[12pt,preprint]{emulateapj}
\usepackage{lscape}

\shorttitle{Survey of NGC2451}
\shortauthors{Balog et al.}

\begin{document}

\title{{\it SPITZER}/IRAC-MIPS Survey of NGC2451A and B: Debris Disks at 50-80 million years}

\author{Zoltan Balog\altaffilmark{1}}
\affil{Steward Observatory, University of Arizona, 933 N. Cherry Av. Tucson, AZ, 85721}
\email{zbalog@as.arizona.edu}

\author{L\'aszl\'o L. Kiss}
\affil{Sydney Institute for Astronomy, School of Physics A28, University of Sydney, NSW 2006, Australia}

\author{J\'ozsef Vink\'o}
\affil{Dept. of Optics and Quantum Electronics, University of Szeged, H-6720, Szeged, Hungary}

\author{ G. H. Rieke, James Muzerolle, Andr\'as G\'asp\'ar,  Erick T. Young}
\affil{Steward Observatory, University of Arizona, 933 N. Cherry Av. Tucson, AZ, 85721}

\author{Nadya Gorlova}
\affil{Department of Astronomy, University of Florida, Gainesville, FL 32611-2055.}

\altaffiltext{1}{on leave from Dept. of Optics and Quantum Electronics, University of Szeged, H-6720, Szeged, Hungary}

\begin{abstract}
We present a Spitzer IRAC and MIPS survey of NGC 2451 A and B, two open clusters in the 50-80 Myr age range. We complement these data with extensive ground-based photometry and spectroscopy to identify the cluster members in the Spitzer survey field. We find only two members with 8 micron excesses. The incidence of excesses at 24 microns is much higher, 11 of 31 solar-like stars and 1 of
7 early-type (A) stars.
This work nearly completes the debris disk surveys with Spitzer of clusters in the 30-130 Myr range. This range is of interest because it is when large planetesimal collisions may have still been relatively common (as indicated by the one that led to the formation of the Moon during this period of the evolution of the Solar System). We review the full set of surveys and find that there are only three possible cases out of about 250 roughly solar-mass stars where very large excesses suggest that such collisions have occurred recently. 
\end{abstract}

\keywords{(Galaxy:) open clusters and associations: individual (NGC 2451); (stars:) circumstellar matter; (stars:) planetary systems: protoplanetary disks; infrared: stars}

\section{Introduction}
Studying the evolution of disks around young stars offers the opportunity to understand the formation of planetary systems and the process that leads to the formation of terrestrial planets. Protoplanetary disks accrete their gas content and evolve into debris disks (produced by the collision of larger sized bodies ) on a timescale of 10 million years. The signatures of debris disks appear as mid-IR excess emission. The fraction of stars displaying mid-IR excess and the magnitude of the excess appears to decrease with time. On top of this overall trend, a few individual stars show extremely large excesses even at several hundred million years. Determining the nature and frequency of the events producing those excesses promises to provide an important perspective on planetary system evolution.

The 30 - 150 Myr range is especially interesting since \citet{Canu04} and \citet{Toub07} showed that the Earth-Moon system was formed by a collision about 30-150 Myr after the Solar System was formed. \citet{Gorl07} called attention to a $\approx$35 Myr old system with a large amount of debris that might have resulted from such a giant collision. \citet{Rhee08, Gorl04, Riek05} and \citet{Su06} also identify young debris systems with large 24$\mu$m excesses. 

\begin{figure*}[!t]
\epsscale{.80}
\plotone{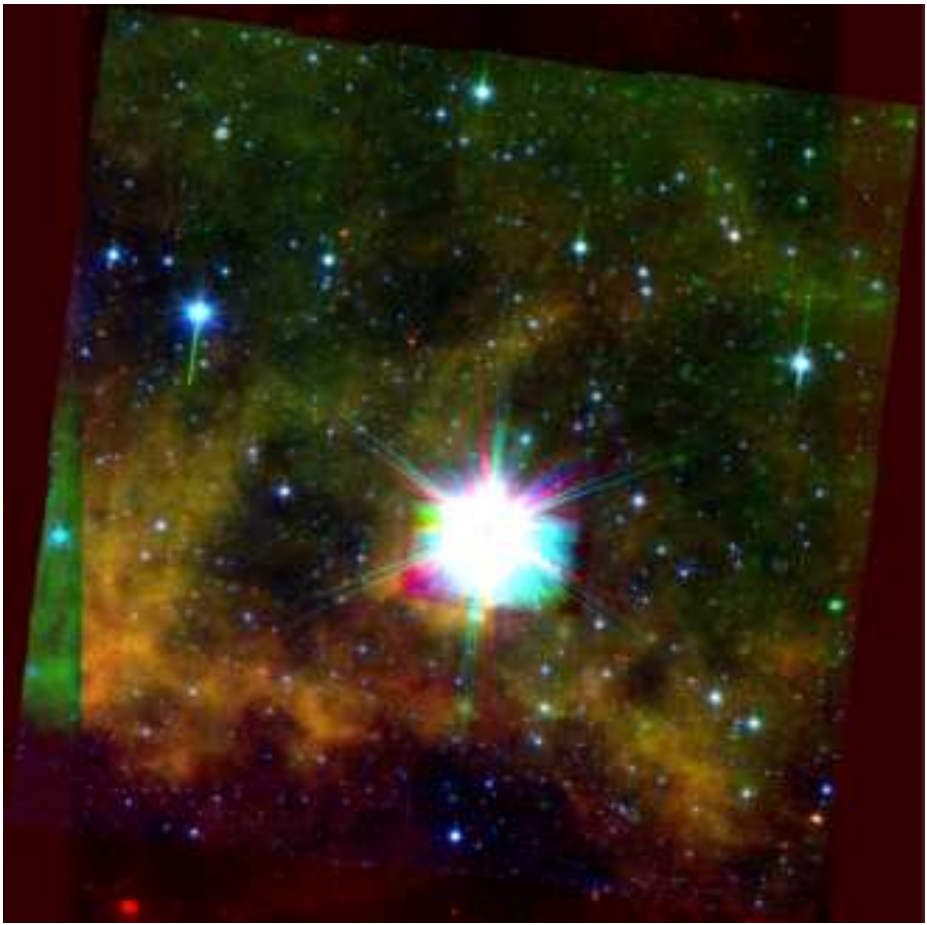}
\caption{False color image of the center of NGC~2451. Blue: 4.5 $\mu$m; Green: 8 $\mu$m; Red: 24 $\mu$m}
\label{fig:RGB}
\end{figure*}

NGC 2451A and B are two young open clusters projected on each other along the same line of sight that are within this interesting age range. Several attempts have been made to separate the two clusters and to determine the physical parameters of each one of them. \citet{Plat01} analysed photometric and spectroscopic data and used proper motions and radial velocities to select members of NGC2451A. They fit theoretical isochrones to the cluster color-magnitude diagram (CMD) to calculate its distance, reddening, and age. They derived d=188 pc, E(B-V)=0.01 and age=60 Myr. \citet{Hunsch03} carried out an X-ray study of the two clusters and identified 39 members of the A and 49 of the B cluster using combined X-ray and optical data. They derived distances of 206 pc and 370 pc, with corresponding ages of 50-80 Myr and 50 Myr for NGC2451A and B, respectively. \citet{Hunsch04} completed the X-ray study with high-resolution spectroscopy and refined the membership of the two clusters. The most recent distance and age estimate was performed by \citet{Kharc05}. They estimated distances of 188 pc and 430 pc and ages of 57.5 Myr and 75.9 Myr for NGC2451 A and B, respectively. However, their analysis was based on only a handful of member candidates (26 and 11 for NGC~2451A and B, respectively) in contrast with \citet{Hunsch03}.  Therefore we adopt the \citet{Hunsch03} distances and ages and reddening (E(B-V) of 0.01 and 0.05 for NGC2451 A and B, respectively) for our analysis.

In this paper we report on a {\it Spitzer} IRAC and MIPS survey of the central region of the clusters to study the incidence of infrared excesses in the 50-80 Myr age range. We supplement our {\it Spitzer} observations with a large-scale spectroscopic and optical photometric program to refine and extend our knowledge of the membership of these clusters. 

\section{Observations}
\subsection{{\it Spitzer}/IRAC and MIPS data}
Observations of NGC2451 were obtained on 2004 March 02 with IRAC \citep{Fazio04}. The IRAC survey covers about 0.6 square degree centered on HD 63032, a K2.5Ib star. The 12s high-dynamic-range mode was used to obtain two frames in each position, one with 0.4s exposure time and one with 10.4s. The observation of each field was repeated twice with a small offset, providing 20.8s integration time for each position. The frames were processed using the Spitzer Science Center (SSC) IRAC Pipeline v14.0, and mosaics were created from the basic calibrated data (BCD) frames using a custom IDL program (see \citet{Guter08} for details). Due to the 7 arcmin offset between channel 1/3 and channel 2/4 the total area covered in all four channels is about 0.45 sq degree. 

Source finding and aperture photometry on these images were carried out using PhotVis version 1.10, which is an IDL-GUI based photometry visualization tool (see \citet{Guter04} for further details on PhotVis). The radii of the source aperture, and of the inner and outer boundaries of the sky annulus were 2.4, 2.4 and 7.2 arc-second, respectively. The calibration used large aperture measurements of standard stars. The zero point magnitudes of the calibration were 19.6642, 18.9276, 16.8468, and 17.3909 corresponding to zero point fluxes of 280.9, 179.7, 115.0,64.13 Jy for channels 1, 2, 3, and 4, respectively. Aperture corrections of 0.21, 0.23, 0.35 and 0.5 mag were applied for channels 1, 2, 3, and 4 to account for the differences between the aperture sizes used for the standard stars and for the NGC~2451 photometry.

We accepted as good detections those with photometric uncertainties less than 0.2 mag, which allowed limiting magnitudes of 18.2, 17.5, 15.4 and 14.8 at 3.6 $\mu$m, 4.5 $\mu$m, 5.8 $\mu$m, 8.0 $\mu$m, respectively. We detected almost 30000 sources at 3.6 $\mu$m, more than 20000 at 4.5 $\mu$m, and over 4000 and 3000 in the 5.8 $\mu$m and 8.0 $\mu$m images. There are more than 2500 sources that were detected in all 4 channels. Based on the 2MASS magnitudes of the detected sources, the limiting mass of our survey is $<$ 0.1 M$_{\odot}$ for NGC2451 A and around 0.15 M$_{\odot}$ for NGC2451 B.

The MIPS \citep{Rieke04} 24 $\mu$m survey was obtained on 2004 Apr. 11. It covers about 0.8 square degrees centered on the same position as the IRAC survey. The observations were carried out in scan map mode. The frames were processed using the MIPS Data Analysis Tool \citep{Gordon05}. We used PhotVis to extract the sources in the 24 $\mu$m image and PSF fitting in the IRAF/DAOPHOT package was used to obtain photometry. We used a 0.59 mag aperture correction and a 7.17 Jy zero point to convert our PSF photometry to the 24 $\mu$m magnitude scale. Similarly to the IRAC photometry we also discarded sources with errors larger than 0.2 mag, leaving 2625 sources in our MIPS sample. The limiting magnitude of our 24 $\mu$m survey is 11.8, which corresponds to masses of 0.55  M$_{\odot}$ (early M spectral type) and 0.9 M$_{\odot}$ (early K spectral type) for stars with pure photospheric emission at the distance of NGC2451 A and B, respectively. 

Fig \ref{fig:RGB} shows the false color image of the center of NGC~2451.

\subsection{Optical photometry}

The optical observations were made with the SITe 2048-\#6 CCD 
camera attached to the 1.5 m telescope at CTIO on 2003 Jan. 22, 23 and 24.
The  camera was mounted at the f/13.5 focal position, covering a 
$15 \times 15$ arcmin$^2$ field-of-view with a resolution of 
0.43 arcsec/pixel for the entire $2048 \times 2048$ pixel$^2$ area. 
The observations were made through Johnson-Cousins $UBVRI$ filters,
from the Tek \#1\footnote{http://www.ctio.noao.edu/instruments/filters} 
filter set. 

The whole cluster was covered by $4 \times 3 =  12$ CCD frames centered
on and around the brightest inner area at R.A. = $07:45:20$, DEC =
$-38:02:00$. One off-cluster area (separated by $\sim 1.5$ deg from 
the cluster center) was also imaged to sample the galactic 
foreground/background object population in the same line-of-sight. 
Each field was imaged three times through the same filter. 
One frame was obtained with a short exposure time  ($10$ s for 
$U$ and $5$ s for $BVRI$) and the other two frames were taken 
with a longer one ($250$ s for $U$, $70$ s for $B$ and $50$ s
for $VRI$). 

The reductions of the raw frames were performed with standard routines 
using {\it IRAF}\footnote{{\it IRAF} is distributed by
NOAO which is operated by the Association of Universities for
Research in Astronomy (AURA) Inc. under cooperative agreement 
with the National Science Foundation}. After trimming the edges of the
frames and subtracting the bias level from each image, 
the frames were divided by a master flat field image
obtained by median combining the available flat field frames for each filter.
Both dome flats and sky flats were taken at the beginning of each night
and combined together into the master flat frames. 
After flat field division, the two long-exposure frames corresponding to the
same filter were averaged to increase the signal-to-noise. 

The photometry of the cluster frames was computed via PSF-fitting
using DAOPHOT implemented in {\it IRAF}. A 2nd order
spatially variable PSF ({\tt varorder=2}) was built for each
frame to compensate for distortions of the PSFs due to
either the optical imaging artifacts in the large field-of-view, or
guiding errors that occured randomly on a few frames. For the
PSF-model, {\tt function=auto} was selected, i.e. each
built-in PSF-function was computed and the best-fitting one was
adopted automatically. In most cases the {\tt penny1} or 
the {\tt penny2} function was the best-fitting one. The
PSF-stars were selected interactively from a sample of the $\sim 50$ 
brightest, non-saturated stars on each frame, omitting the ones 
with suspicious profiles and/or detectable neighbors within $r = 15$ pixels.
The {\tt fitrad} parameter was set as $5$ pixels corresponding
roughly to the FWHM of the frames with the lowest quality. 
The detection threshold was fixed at the $4 \sigma$ level on
each frame. 

Inspecting the results of the PSF photometry revealed that there is
a small, but significant, difference between the results of the PSF  and 
aperture photometry of the same stellar field. This difference was 
about the same (within $\sim 0.03$ mag) for the bright stars in one field, 
but its amount varied from field to field, ranging from a few hundredths up to
$\sim 0.2$ mag. The aperture photometry was computed with 
$r_{ap} = 8$ pixels aperture radius (about 3.4$\arcsec$). The local sky level was estimated
as the mode of the pixel distribution within an annulus having inner 
and outer radii of $10$ and $20$ pixels, respectively, centered on each 
object.  Because $r_{ap}$ was more than twice the stellar FWHM for
most frames, much of this difference was probably due to the 
distorted stellar PSFs mentioned above, plus the problems in
determining the local sky level during the PSF-fitting. Therefore, we fixed the zero point of the photometry by selecting a bright,
unsaturated, uncrowded reference star on each frame, and using its
magnitude from aperture photometry as a reference level. Thus, the
final instrumental magnitudes were determined as follows: first, 
differential magnitudes were computed between the PSF-magnitudes
of all stars and the reference object, then the magnitude of the reference
object from aperture photometry was added to each differential magnitude.
For the short-exposure frames, the same procedure was applied, but 
for the reference object the magnitudes from the long-exposure frames were selected,
thus ensuring a common zero point for the photometry of the 
long- and short-exposure frames.

The transformation of the CTIO instrumental magnitudes into the standard
Johnson-Cousins system was performed via observations of Landolt 
photometric standard sequences \citep{Land92}. 
The L98, L101 and RU152 standard fields were imaged through each filter, 
weather and time permitting, on each night at different airmasses (up to 4 different
elevations in the case of the L98 field). Since photometric accuracy is essential
for the conclusions of this paper, we give the details of our calibration and
standard transformation below.

The following formulae were adopted for converting the instrumental magnitudes
into standard ones:
\begin{equation}
V - (v - k_v X) ~=~ \alpha_{CI} \cdot CI ~+~ \zeta_V, 
\end{equation}
\begin{equation}
CI ~=~ \beta_{CI} \cdot (ci - k_{ci} X) ~+~ \zeta_{CI}, 
\end{equation}
where $v$ and $ci$ are the instrumental $V$-magnitude and color index
($u-b$, $b-v$, $v-r$ or $v-i$), $V$ and $CI$ are the standard magnitudes
and colors, $X$ the airmass, the $k$s are the extinction coefficients, 
$\alpha_{CI}$ and $\beta_{CI}$ are the color-dependent 
transformation slopes (the color terms), and the $\zeta$s are the zero points. 

\begin{deluxetable}{lccccc}
\tablecaption{\label{tbl-stdtrafo}Transformation coefficients derived from Landolt standards observed on 2003 Jan.22.}
\tablehead{
\colhead{Coefficient} & \colhead{$V$} & \colhead{$(U-B)$} & \colhead{$(B-V)$} & \colhead{$(V-R)$} &\colhead{ $(V-I)$}
}
\startdata
$k$ & 0.15 & 0.15 & 0.13 & 0.00 & 0.05 \\
$\alpha$ & -- & -- & 0.035 & -- & 0.031 \\
$\beta$ & -- & 1.064 & 0.880 & 1.055 & 1.021 \\
$\zeta$ & $-2.457$ & $-1.848$ & $-0.133$ & $-0.083$ & $0.822$ \\
\enddata
\end{deluxetable}

Because the first night was nearly photometric, we decided to determine
the color terms and the extinction coefficients from the
magnitudes of standard stars observed that night. The instrumental
magnitudes of these stars were computed from aperture photometry
using the same aperture and annulus radii as mentioned above for the cluster 
frames.  The transformation coefficients were then derived by fitting
Eq.1 and 2 to the instrumental and standard magnitudes via 
$\chi^2$-minimization. Table~\ref{tbl-stdtrafo} lists the parameters 
for each filter combination (note that both $(B-V)$ and $(V-I)$
were applied as the color index in Eq.1).

With the parameters listed in Table \ref{tbl-stdtrafo} determined, the
color terms and extinction coefficients were kept fixed and only the
zero points were fitted to the data for standard stars observed on the
next two nights. Because we experienced slightly variable transparency 
during the second night, we have not attempted to adjust the
extinction correction coefficients to achieve the best fit. 
The variation of the zero points (collected together in Table \ref{tbl-zpt})
is under $0.01$ mag except for $(U-B)$, illustrating the stability
of the photometric conditions at the CTIO site during our observing run.

\begin{deluxetable}{lccccc}
\tablecolumns{6}
\tablewidth{0pc}
\tablecaption{\label{tbl-zpt}Transformation zero points for the three nights}
\tablehead{
\colhead{Date}& \colhead{$\zeta_{V}$} & \colhead{$\zeta_{U-B}$} & \colhead{$\zeta_{B-V}$} & \colhead{$\zeta_{V-R}$} & \colhead{$\zeta_{V-I}$}
}
\startdata
2003 Jan 22 & $-2.457$ & $-1.848$ & $-0.133$ & $-0.083$ & $0.822$ \\
2003 Jan 23 & $-2.470$ & $-1.770$ & $-0.133$ & $-0.083$ & $0.800$ \\
2003 Jan 24 & $-2.486$ & $-1.768$ & $-0.134$ & $-0.082$ & $0.802$ \\
\enddata
\end{deluxetable}

Inserting the coefficients from Table \ref{tbl-stdtrafo} and \ref{tbl-zpt}
into Eq.1 and 2, the photometric data for the stars in the cluster fields 
have been transformed into the standard system. This resulted in
a final sample of $\sim 29,000$ stars having calibrated photometry in
at least the 

\begin{figure}[b]
\plotone{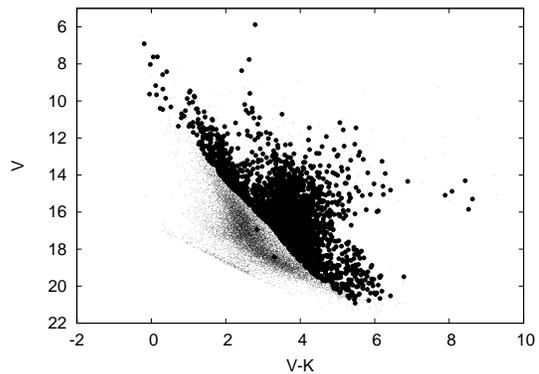}
\caption{Spectroscopic sample on the V vs V-K diagram. Large dots: targets with spectra, small dots: all stars with available V and K magnitudes in the whole cluster area}
\label{fig:specsampl}
\end{figure}$V$ and $I$ filters with uncertainties less than $0.1$ magnitude.

The quality and the stability of the whole photometry 
including the standard transformation
has been verified by comparing our standard V and B-V magnitudes with those
from \citet{Plat01} for the $\sim 1800$ stars in common between the two
samples. The agreement is
very good. The mean difference between the two samples is $0.001 \pm 0.06$ mag in $V$ and $-0.007 \pm 0.04$ mag in
$B-V$ (the errors are the RMS of the residuals), thus, statistically insignificant. Therefore
our new CCD photometry in the field of NGC~2451 results in
a sample containing $\sim 29,000$ stars with calibrated data.
This nearly doubles the amount of photometric data available 
for this field.

\subsection{Radial velocity survey}

We acquired AAOmega data using the Anglo-Australian Telescope at Siding Spring,
Australia on three nights, February 15, 16 and 17, 2008, in acceptable conditions (clear skies with 1.5-2.5 arcsec seeing). In the blue arm we
used the 2500V grating, providing $\lambda/\Delta\lambda=8000$ spectra between
4800 \AA\ and 5150 \AA. In the red arm we used the 1700D grating that has been
optimized for recording the Ca II infrared triplet region. The red spectra range
from 8350 \AA\ to 8790 \AA, with $\lambda/\Delta\lambda=10000$. This setup has
the highest spectral resolution available with AAOmega, suitable to measure stellar radial velocities. In total, we acquired eleven field
configurations centered on the open clusters. We selected the targets for our
spectroscopic campaign based on their positions in the optical--near-IR
color-magnitude diagram: we targeted each source that is brighter than the 80
Myr isochrone plus 0.2 mag placed at the distance of the more distant cluster in
the V vs. V--K diagram. We show the spectroscopic candidates in Fig. \ref{fig:specsampl} together with all the stars with available V and K photometry in the field. Our spectroscopic coverage is not complete for two reasons: (i) we concentrated our efforts to the central part of the cluster where we had $Spitzer$ data; (ii) there were faint objects for which the low S/N of the spectra prevented measuring meaningful radial velocities.

\begin{figure}
\begin{center}
\leavevmode
\includegraphics[width=8cm]{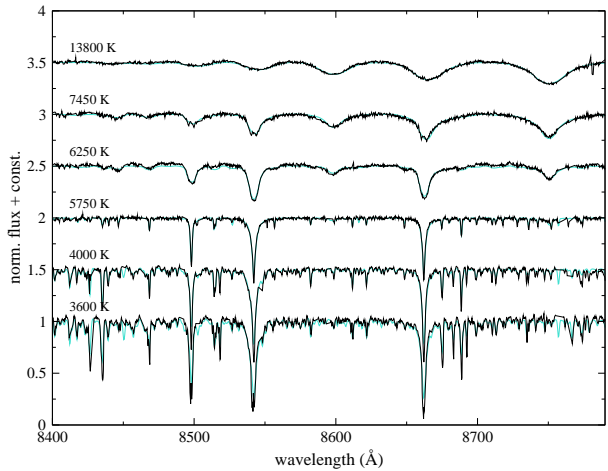}
\caption[]{\label{specfit} Observed stellar spectra (light blue/gray lines) and
the best-fit synthetic data from the \citet{Munar05} spectrum library (black lines).}
\end{center}
\end{figure}

\begin{figure*}[!t]
\plottwo{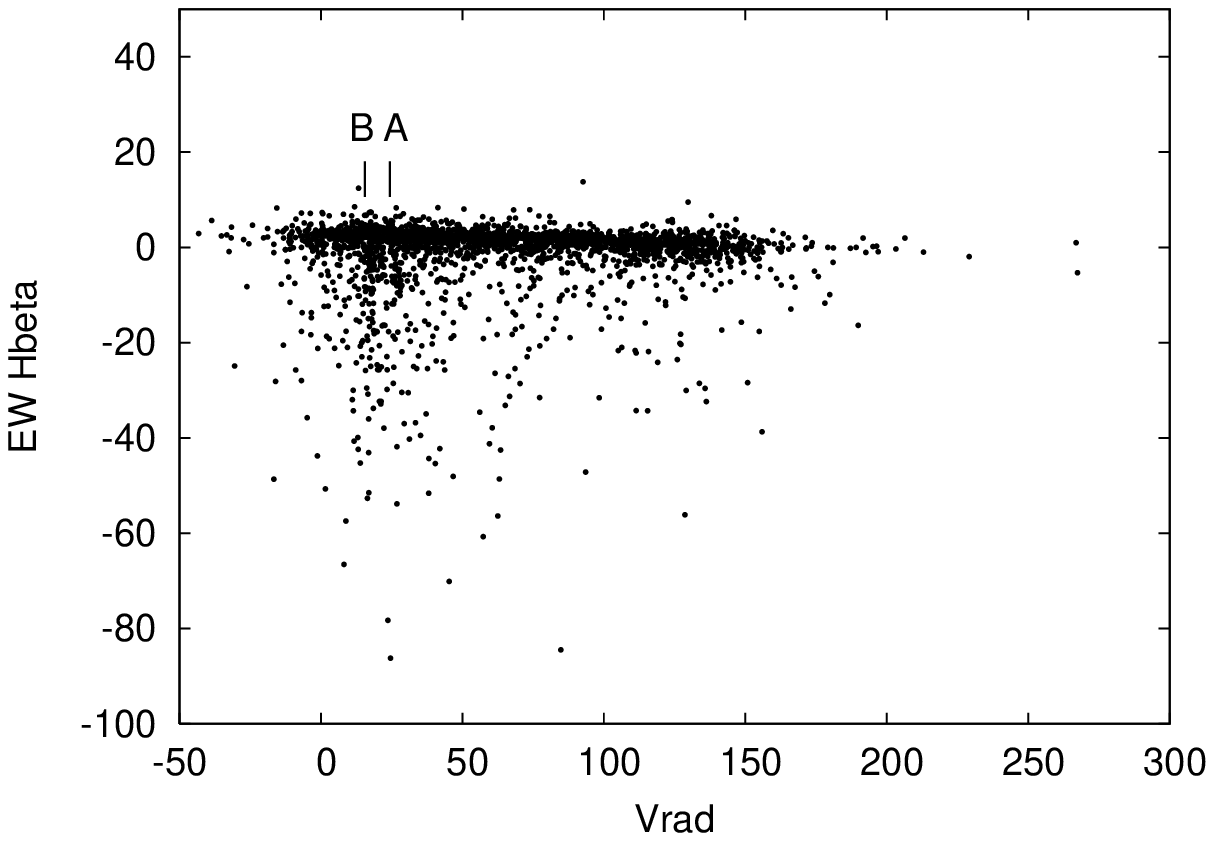}{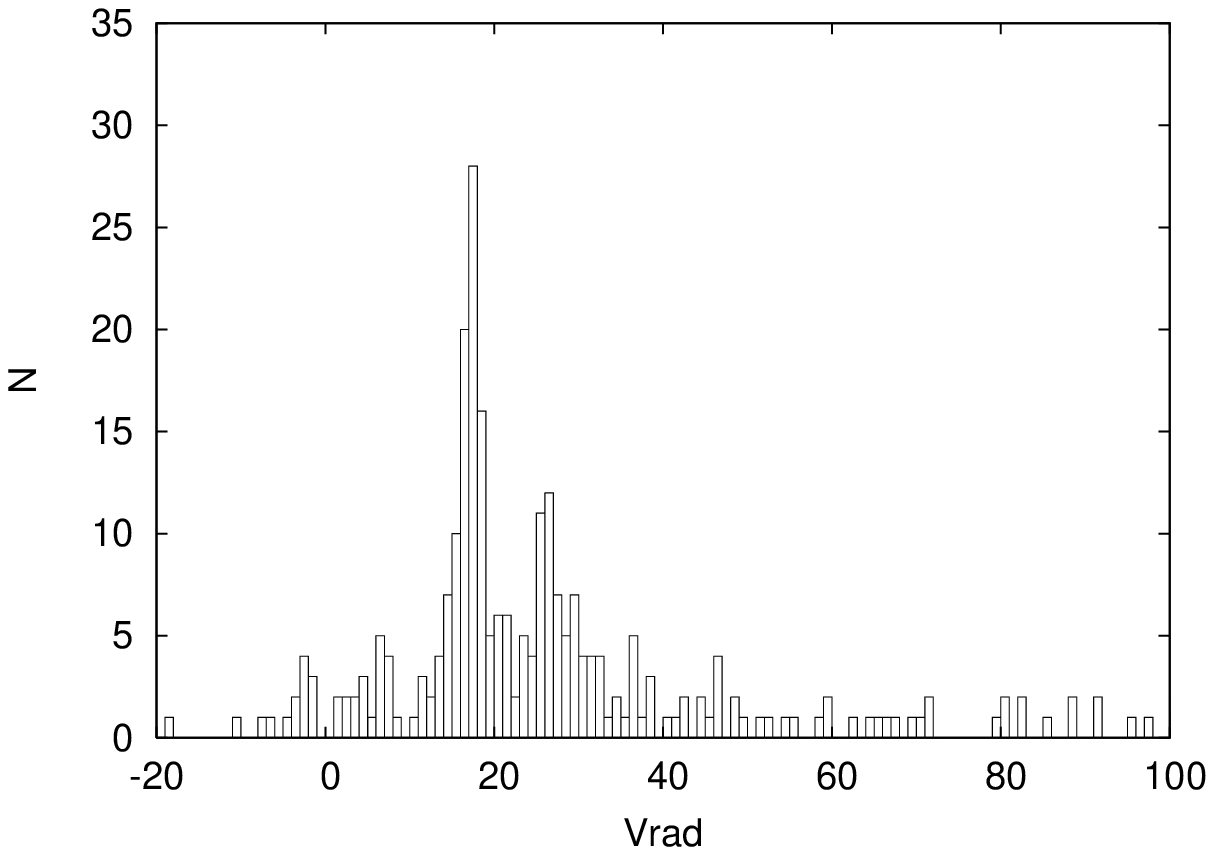}
\caption{Left panel: EW of H$\beta$ vs radial velocity. A and B show the clusters' mean radial velocities from \citet{Hunsch04}. Left panel: all stars, right panel: radial velocity histogram of the two clusters using stars that show evidence of chromospheric activity}
\label{fig:radvel_hist}
\end{figure*}

The spectra were reduced using the standard 2dF data reduction pipeline. We
performed continuum normalization for the stellar spectra using the IRAF task
onedspec.continuum and then cleaned the strongest skyline residuals using linear interpolation of the surrounding continuum.

Atmospheric parameters and radial velocity were determined for each star in an iterative process, which combined finding best-fit synthetic spectrum from the \citet{Munar05} spectrum library, with $\chi^2$ fitting, and cross-correlating the best-fit model with the observed spectrum to calculate the
radial velocity. This approach is very similar to that adopted by the Radial
Velocity Experiment (RAVE) project \citep{Stein06,Zwitt08},
and our analysis is based on the same synthetic library as RAVE. Earlier results for other star clusters with the same instrument and analysis can be found in \citet{Kiss07,Kiss08}. Because of the wide range of temperatures (and hence spectral features), we needed three subsequent iterations to converge to a stable set of temperatures, surface gravities, metallicities and radial velocities. The latter are believed to be accurate within $\pm$1-2 km~s$^{-1}$ for the cooler stars and $\pm$5 km~s$^{-1}$ for the hotter stars in the sample (the boundary is roughly at 8000-9000 K). These values have been estimated from Gaussian fits of the cross-correlation profile using the IRAF task rv.fxcor and should only be considered as representative numbers. The atmospheric parameters ($T_{\rm eff}$, $[M/H]$, $\log~g$), due to their degeneracy, are accurate to about $\pm$300 K and $\pm$0.5 dex, with strong correlations (see \citet{Zwitt08} for a thorough discussion of the RAVE error analysis).

To illustrate the difficulties one faces when analysing cool and hot stars
together in the Ca~II triplet region, we show sample spectral fits in Fig. \ref{specfit}. Since the Ca~II lines almost exactly coincide with hydrogen lines in the Paschen series, we found that it was absolutely crucial to have the best-match template for cross-correlation. A slight template mismatch can easily lead to radial velocity shifts of several km~s$^{-1}$ at this intermediate spectral resolution and hence one has to be very careful to optimize template selection (contrary to the commonly used practice in the optical range that the same template is used across a range of spectral subtypes or even types). It is also unavoidable that as soon as the temperature reaches about 9000 K, the broad spectral features will lead to a degraded velocity precision simply because of the broadened cross-correlation profile. NGC~2451A and B, as reasonably young open clusters, still host a significant number of hotter main sequence stars, possibly resulting in degraded velocity precision for a fraction of stars.

Another difficulty was the very high incidence of Ca~II emission (defined by the excess flux of the calcium lines relative to the best-fit model spectra), especially in late-type main sequence stars in the clusters. Both
clusters are young, hence the high rotation rates can lead to elevated chromospheric activity, for which the Ca~II lines are good indicators \citep{Andr05}. When comparing the best-fit models to the observed spectra, we noticed distorted Ca~II profiles for many stars, frequently in the form of central emission components within the lines. These can introduce random velocity errors of up to 10-20 km~s$^{-1}$ in the maximum of the cross-correlation profile. This was particularly apparent after modifying the velocity determination for 405 stars with Ca~II line profile irregularities: after excluding the Ca~II lines from cross-correlations, a large fraction of the refined velocities were very close to the cluster's mean velocities (Fig.\ref{fig:radvel_hist}).

We also measured cross-correlation velocities from the lower resolution blue spectra. These data were less useful because many of the stars have a broad H$\beta$ line and only a few additional weak features in the recorded wavelength range, leading to a broad cross-correlation profile and radial velocities accurate to only about $\pm$10 km~s$^{-1}$.

\section{Analysis}
\subsection{Radial velocities of the clusters}

The clusters are in the plane of the Milky Way so the background contamination overwhelms the cluster members. To estimate the radial velocities of the clusters we need to make some assumptions to separate the probable cluster members from the field based on our present knowledge of the cluster. Based on the young age of the clusters, we assumed that our sample should contain a large number of K and M dwarfs  that show chromospheric activity, i.e., signs of emission in the CaT lines. We identified 405 active red objects in our sample. Several of our target stars show sharp H$\beta$ emission as additional evidence for activity. In the left panel of Fig. \ref{fig:radvel_hist} we show the equivalent width of H$\beta$ vs radial velocity (negative EWs mean emission). We see two distinct groups of emission line objects in the figures around the approximate position of the radial velocity values of NGC2451 A and B (22.7 ${\rm km s^{-1}}$ and 14 ${\rm km s^{-1}}$ ) that were estimated by \citet{Hunsch04}. We show the radial velocity histogram of these stars in the right panel of Fig. \ref{fig:radvel_hist}. From this histogram we calculate mean radial velocities 26.7$\pm 2.4$  ${\rm km s^{-1}}$  and 17.4$\pm 2$ ${\rm km s^{-1}}$ for NGC2451 A and B, respectively. Because chromospheric activity is not always present in young stars (e.g. early type ones) we use this test only to estimate true cluster velocities. 

\begin{figure}[t]
\plotone{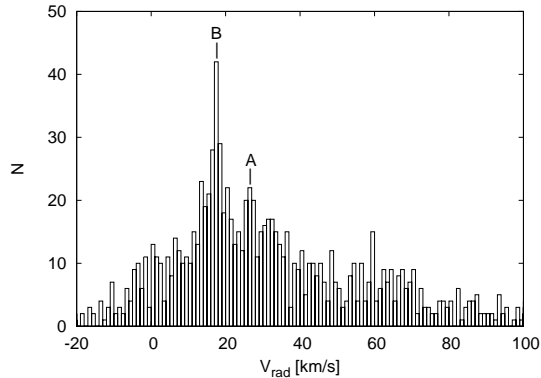}
\caption{Radial velocity histogram of the two clusters using all stars with log g $>$ 3.0. A and B show the radial velocity values determined using the chromospherically active young stars for NGC 2451 A and B, respectively.}
\label{fig:radvel_hist_all}
\end{figure}

\begin{figure*}[!t]
\begin{center}
\includegraphics[scale=0.5]{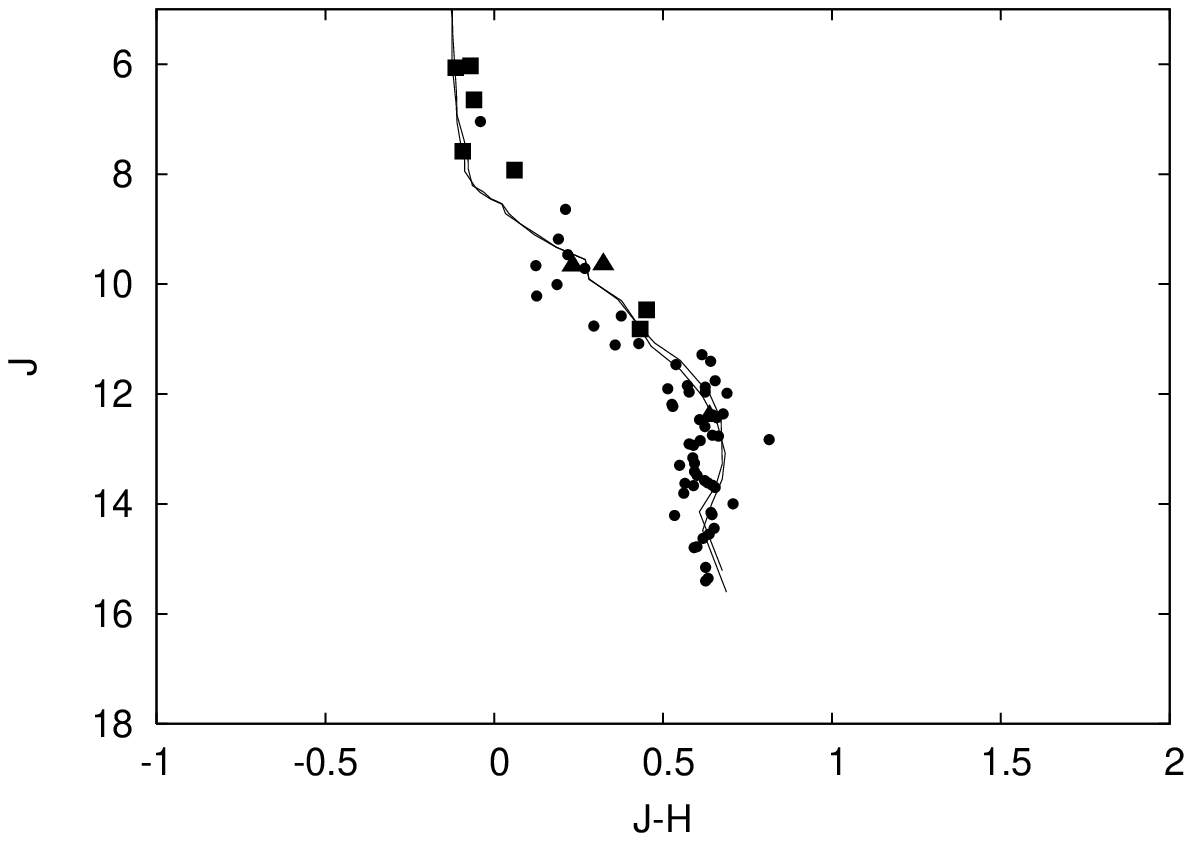}
\includegraphics[scale=0.5]{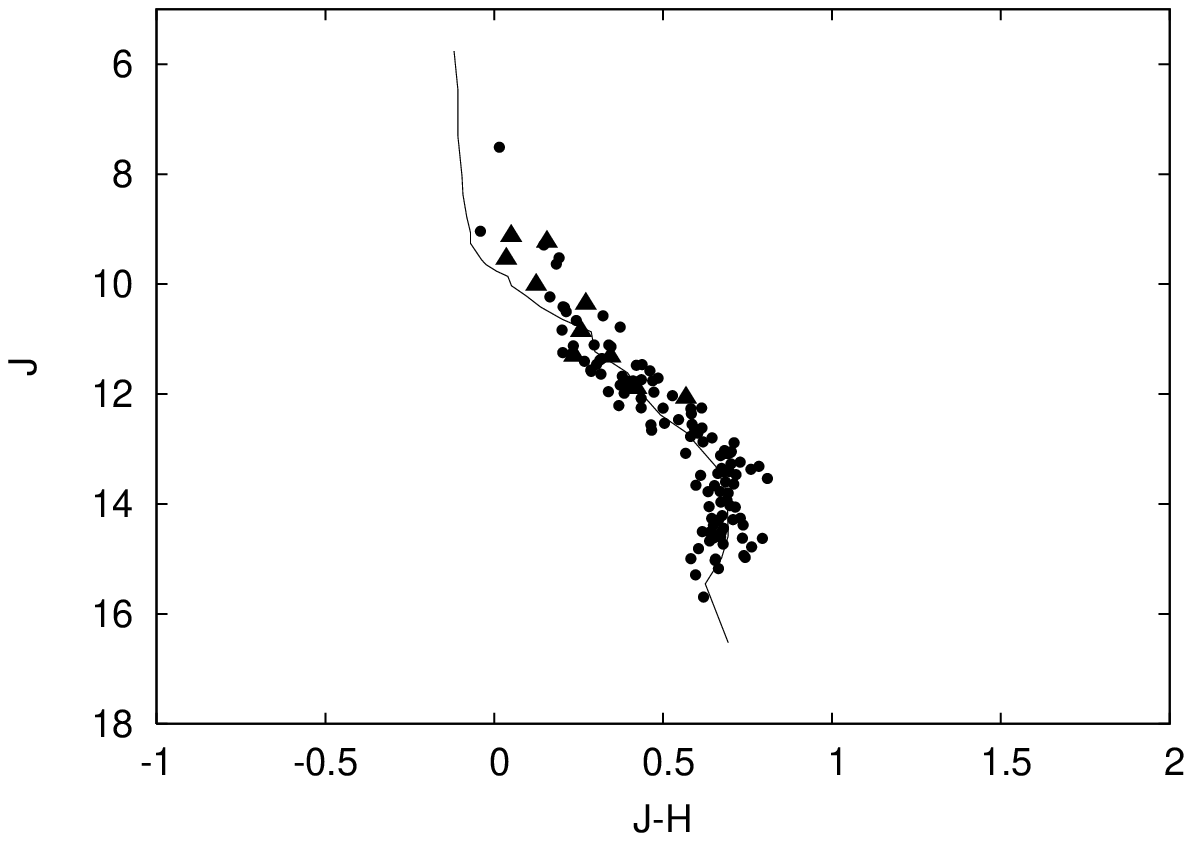}

\includegraphics[scale=0.5]{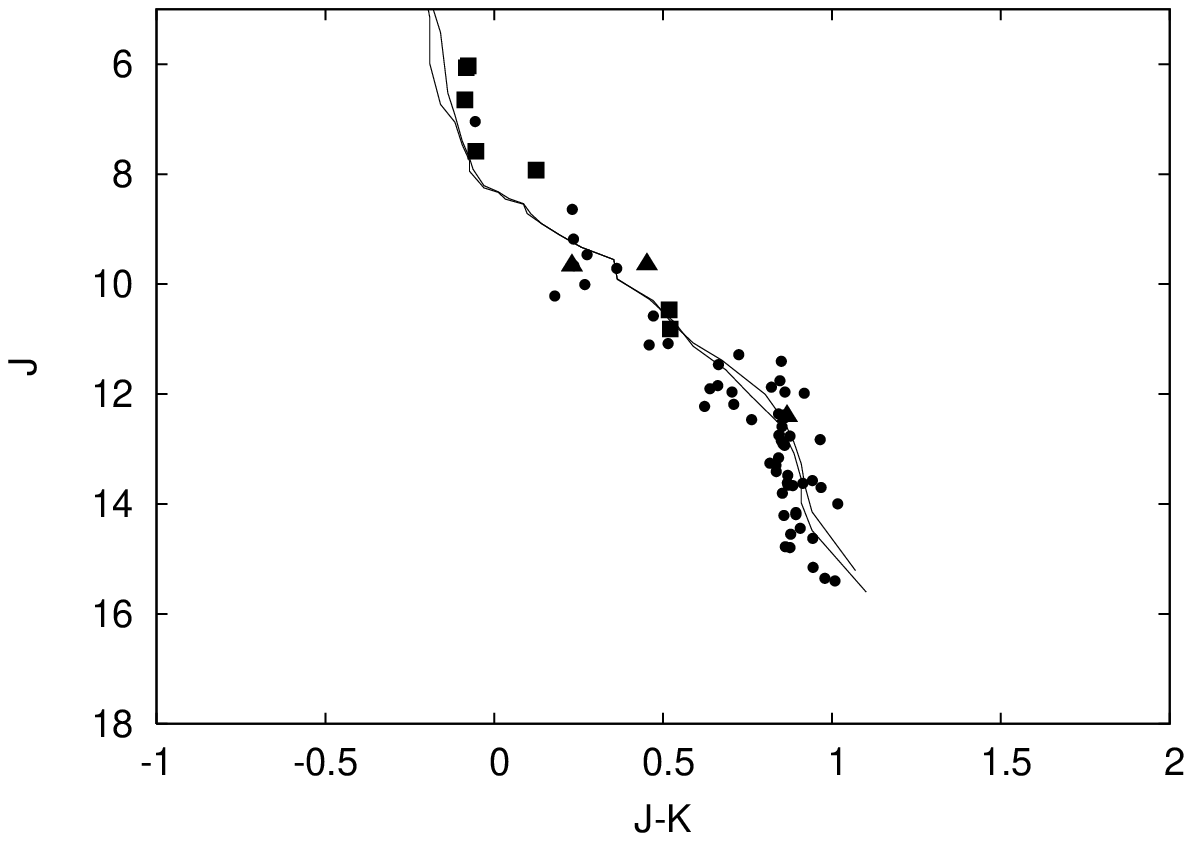}
\includegraphics[scale=0.5]{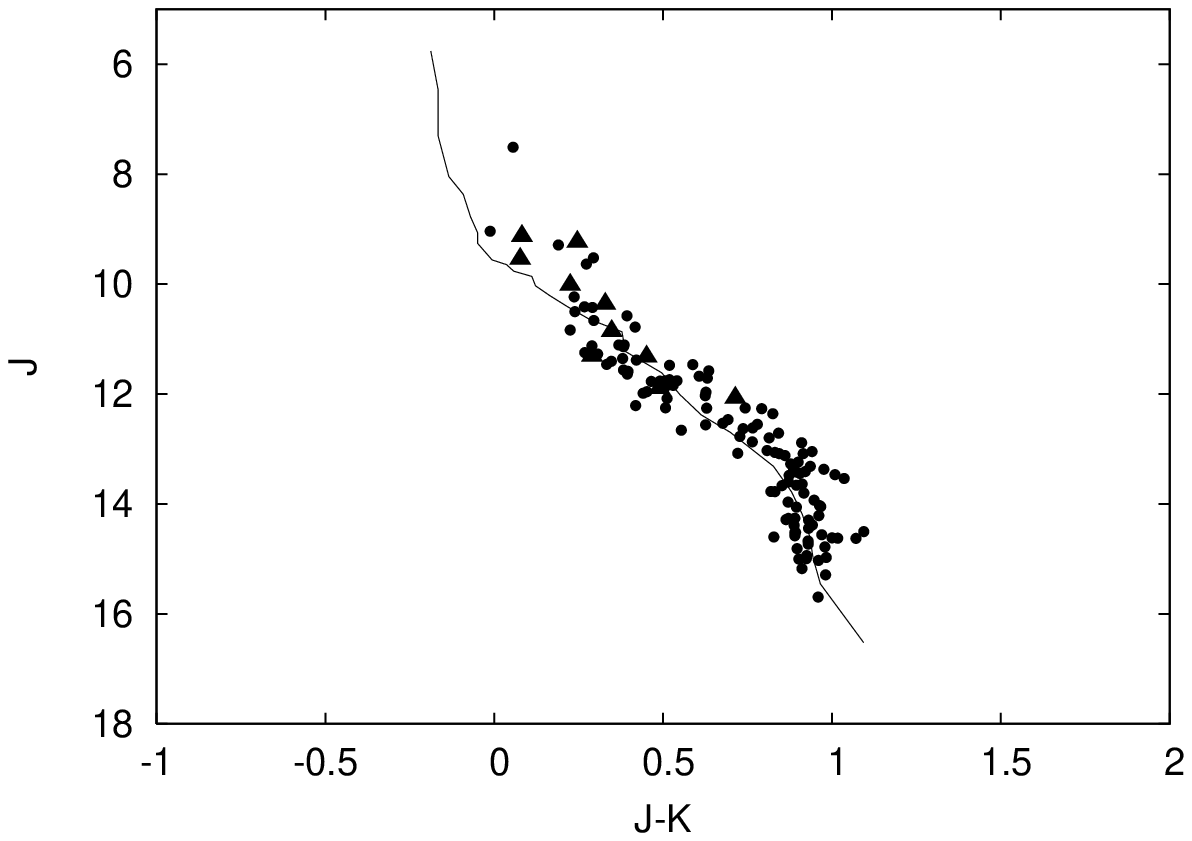}

\includegraphics[scale=0.5]{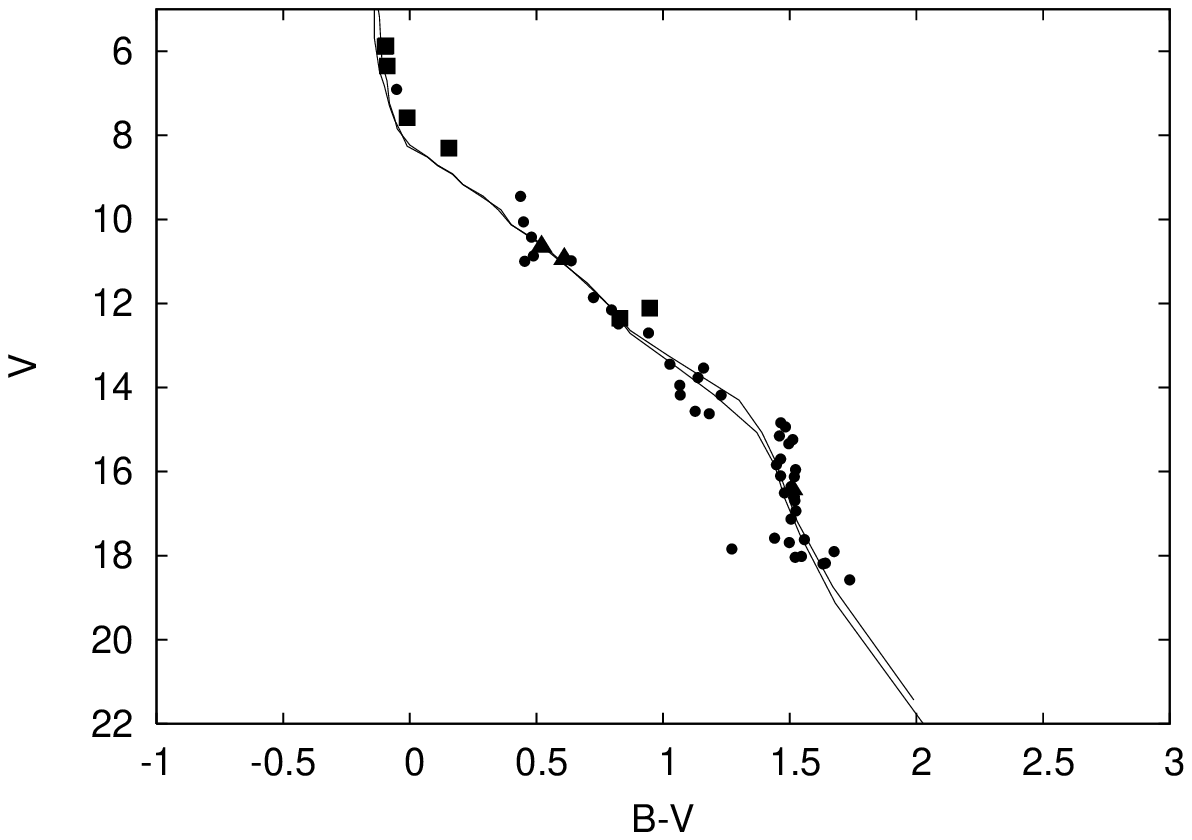}
\includegraphics[scale=0.5]{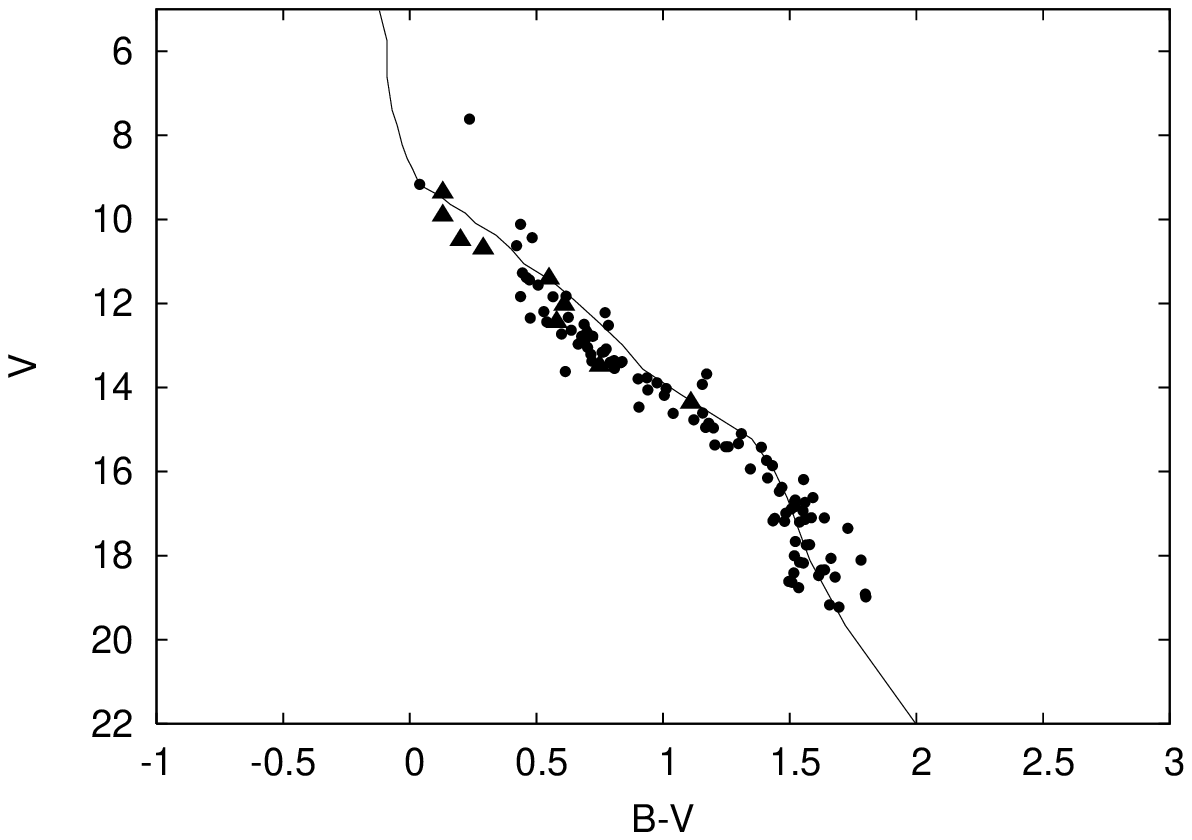}

\includegraphics[scale=0.5]{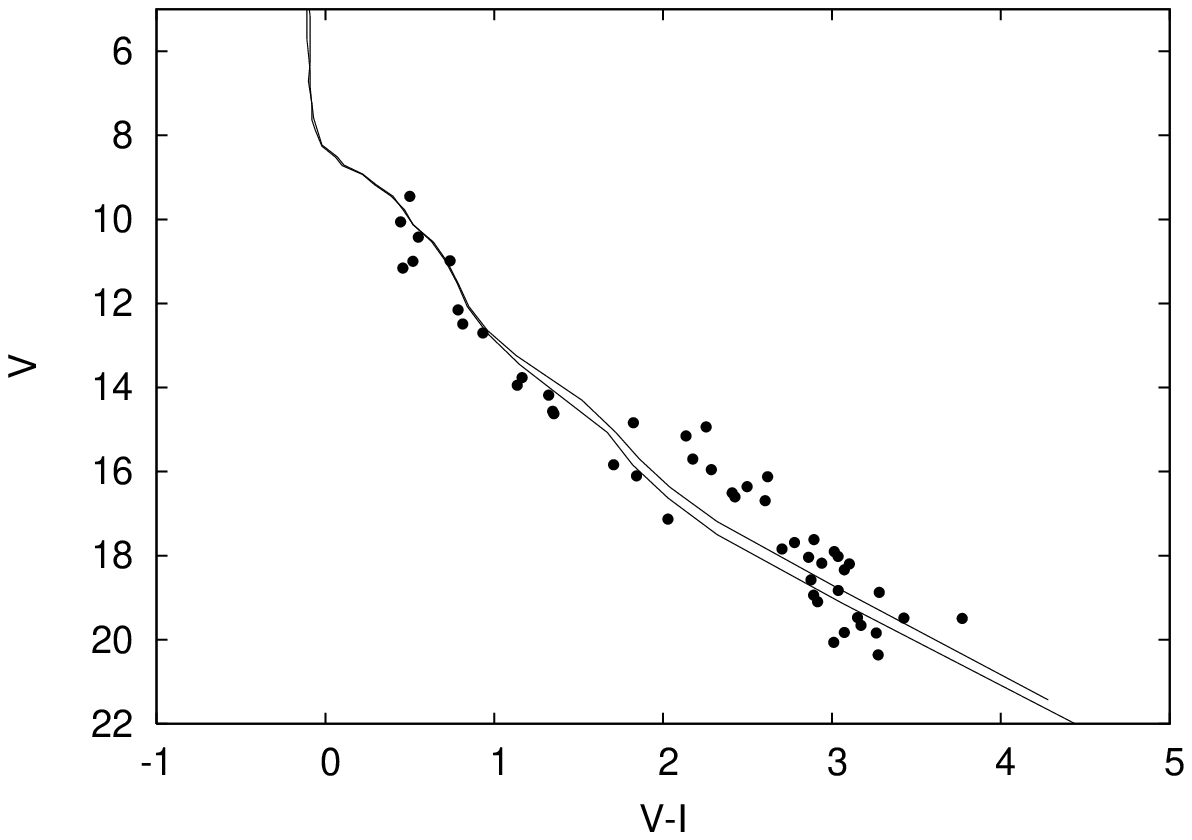}
\includegraphics[scale=0.5]{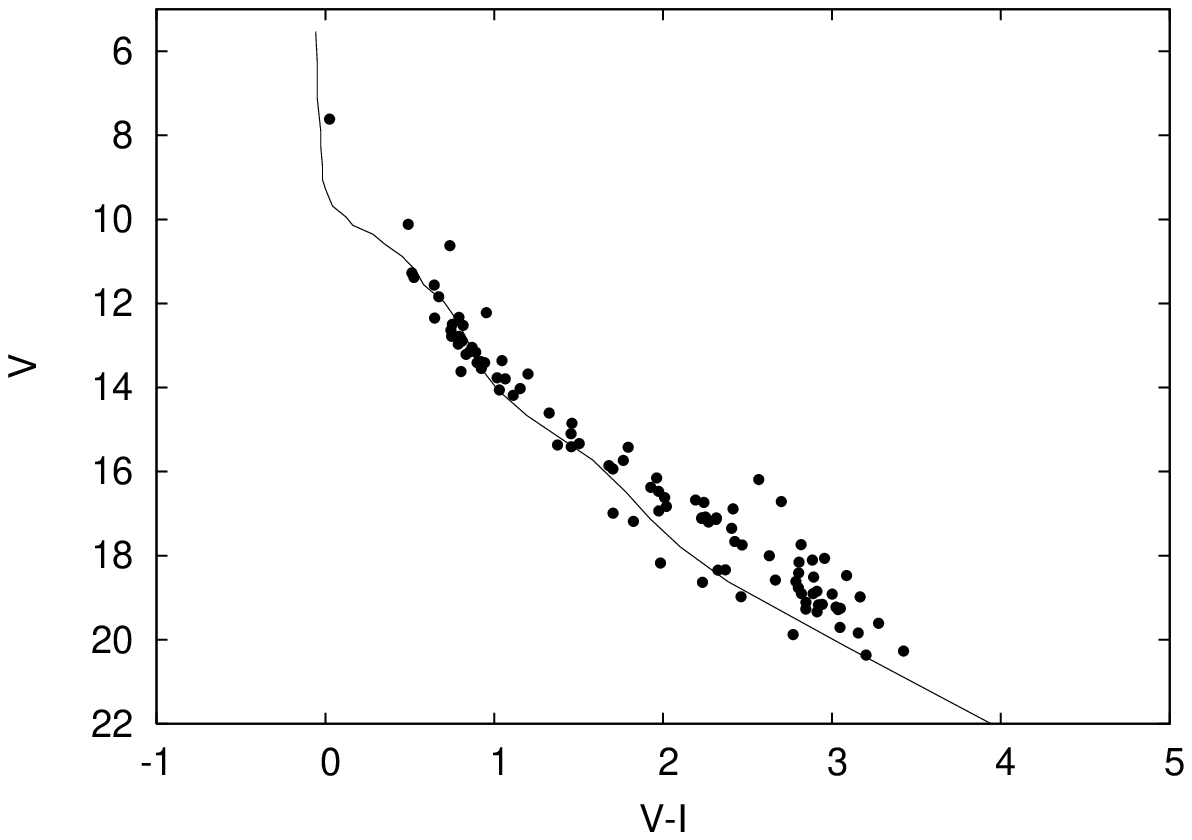}

\includegraphics[scale=0.5]{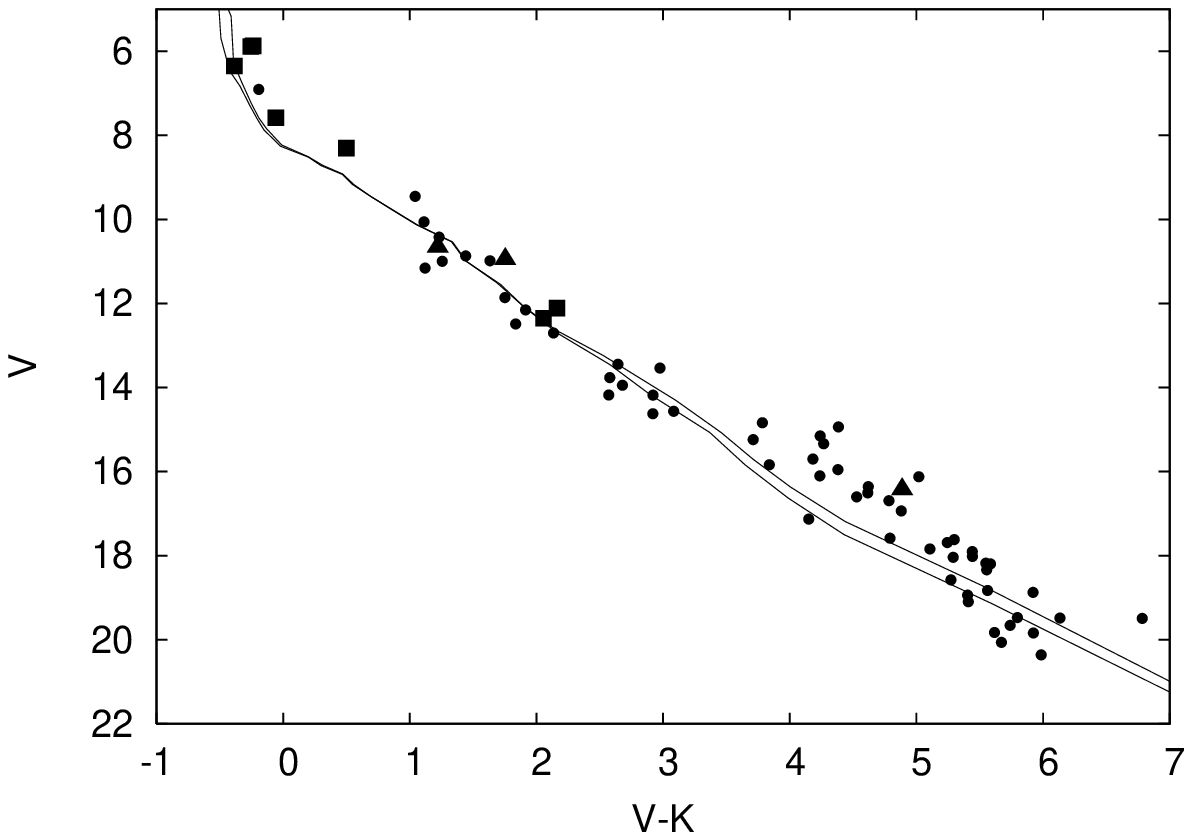}
\includegraphics[scale=0.5]{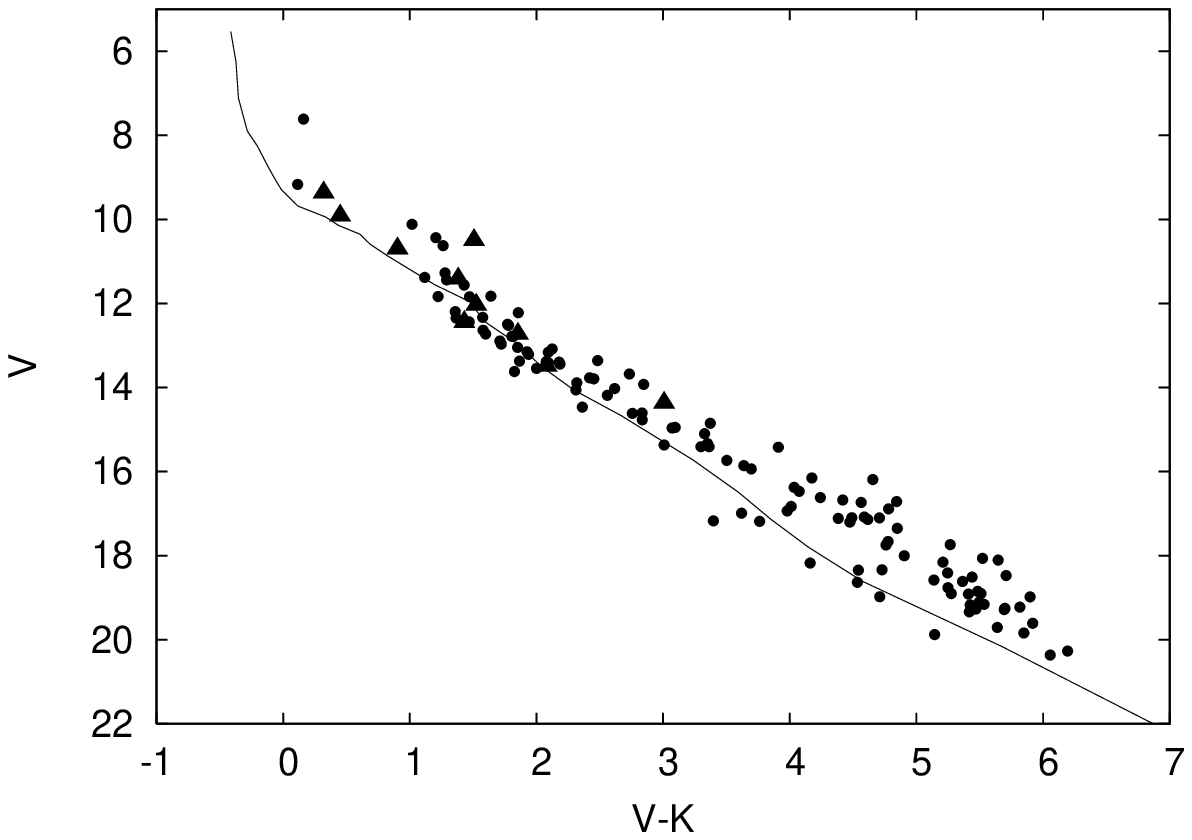}
\caption{Color-magnitude diagram of the members of the two clusters (left panel: NGC2451 A; right panel: NGC2451 B). dots: stars in our spectroscopic survey, squares: members from \citet{Plat01} not included in our spectroscopic sample, triangles: members from \citet{Hunsch03} not included in our spectroscopic sample. The solid line show the 50Myr and 80Myr isochrone of \citet{Siess00} for NGC2451 A and the 50 Myr isochrone of \citet{Siess00} for NGC2451 B, shifted with the distance and extinction of \citet{Hunsch03}}
\label{fig:cmds}
\end{center}
\end{figure*}

Our average radial velocity values are about 3 km~s$^{-1}$ higher than those of \citet{Hunsch04}. Considering the formal standard deviation of the  \citet{Hunsch04}  data (2.3 km~s$^{-1}$ for cluster A and 2.5 km~s$^{-1}$ for cluster B) and the small number of member stars in that sample (21 for cluster A and 10 for
cluster B), the 3 km~s$^{-1}$ systematic difference is hardly significant. Given that our sample contains three-to-ten times more members for each cluster, we
 adopt our mean values for the cluster velocities.

\subsection{Cluster membership and color magnitude diagrams}

In Fig. \ref{fig:radvel_hist_all} we show the radial velocity histogram of our complete sample with log g $>$ 3 (giants filtered out). The two peaks at the positions of the radial velocities of the two clusters (designated with A and B) are readily distinguishable. However a large number of contaminating stars might be present due to the strong galactic background so the radial velocity is not enough to determine accurate membership status. Therefore, we used photometric diagrams also to clean our initial sample. 

\begin{figure*}[!t]
\begin{center}
\includegraphics[scale=0.6]{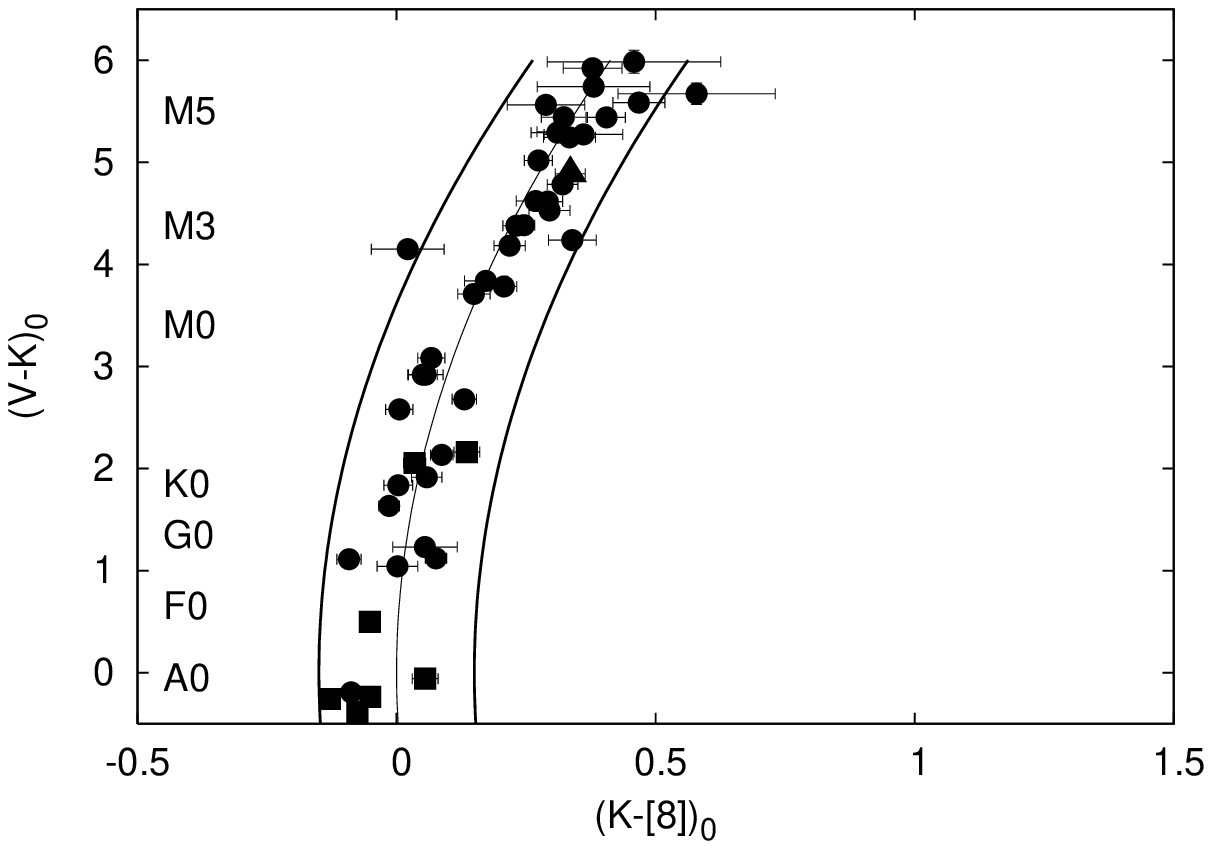}
\includegraphics[scale=0.6]{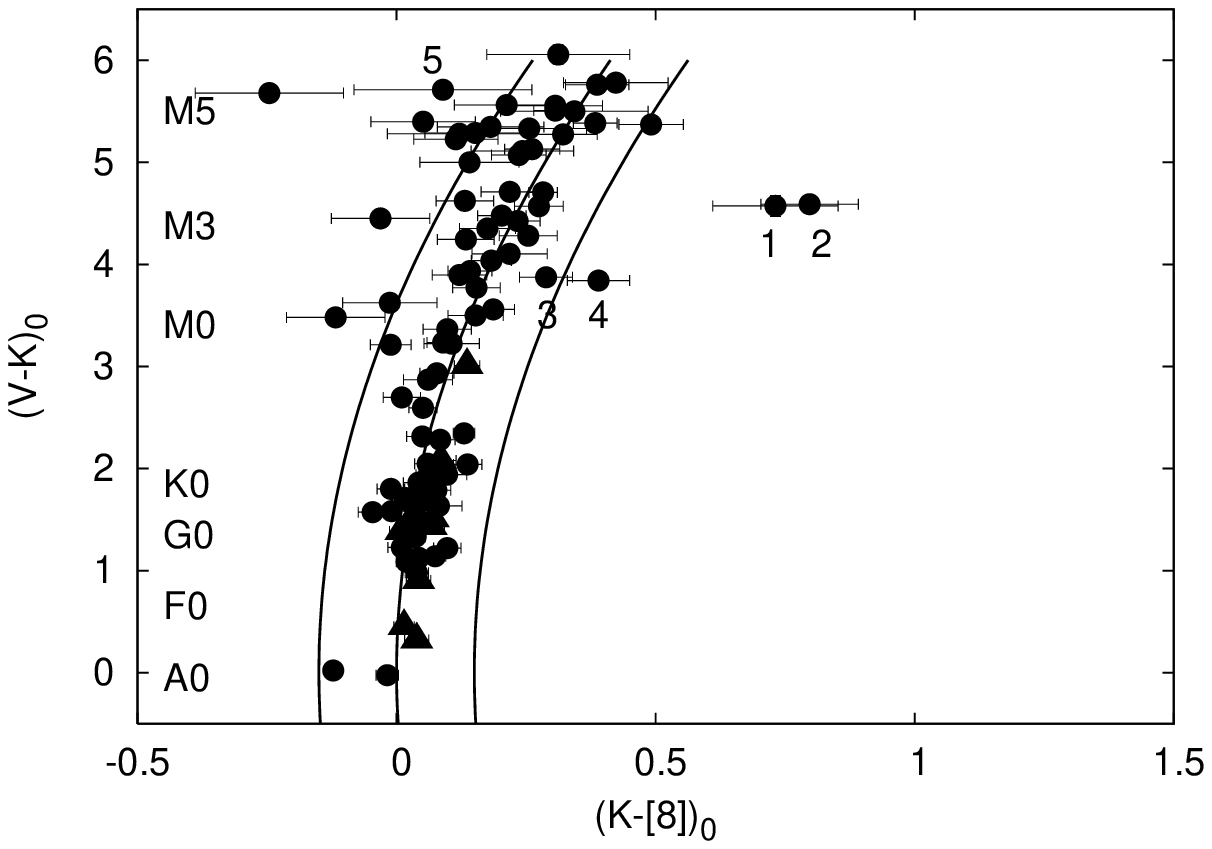}
\caption{V-K vs K-[8] diagrams of NGC 2451 A (left panel) and B (right panel). The meaning of the symbols is the same as in Fig \ref{fig:cmds}. Horizontal error bars show the 1 $\sigma$ photometric errors. The solid lines represent the color-color relation for normal stars and its 3 $\sigma$ limits.}
\label{fig:ccds}
\end{center}
\end{figure*}

We performed the final determination of memberships as follows. We selected member candidates using the radial velocity and log g criteria. We accepted all objects as cluster members that had log g $>$ 3 (1079 out of 2570 objects satisfied this criterion) and not more than a 2 $\sigma$ deviation from the clusters' mean radial velocities (13.4 ${\rm km s^{-1}}$ $<$ V$_{rad}$ $<$ 21.4 ${\rm km s^{-1}}$ for NGC2451 B and 21.9 ${\rm km s^{-1}}$ $<$ V$_{rad}$ $<$ 31.5 ${\rm km s^{-1}}$ for NGC2451 A, 152 and 200 objects for NGC2451 A and B, respectively). Then we plotted the initial sample in color magnitude diagrams and removed the objects with positions inconsistent with the 50 Myr isochrone of \citet{Siess00} at the distance of the clusters. We show the final cleaned diagrams in Fig \ref{fig:cmds}. Our final sample consists of 60 objects in NGC2451 A and 121 objects in NGC2451 B The photometry of the  members is listed in Tables \ref{tab:membAubvri}, \ref{tab:membBubvri}, \ref{tab:membAspitz}, and \ref{tab:membBspitz}.

\subsubsection{Members from \citet{Plat01}}

There are 18 members of NGC2451 A reported by \citet{Plat01} in the area covered by all 5 bands of our {\it Spitzer} observations. Thirteen of them were included in our radial velocity survey. We rejected 7 of these stars based on their radial velocity; 4 were included in our clean NGC2451 A membership list while 2 have radial velocities just outside the 2 $\sigma$ limit. We included these two stars along with the other 5 not covered by the radial velocity survey in our final sample for NGC2451 A (they are marked with filled squares in the plots in the left column of Fig \ref{fig:cmds}). We used the \citet{Plat01} magnitudes for these objects because they were not detected by our UBVRI survey. As a result, they are not plotted in the V vs V-I plot in Fig \ref{fig:cmds}

\subsubsection{Members from \citet{Hunsch03} and \citet{Hunsch04}}

\citet{Hunsch03} list 87 objects in the field of our {\it Spitzer} survey as candidate members of either NGC2451 A or B. Unfortunately the coordinates given in \citet{Hunsch03} are from X-ray data and have sufficiently large uncertainties that it is not easy to identify their sources in the IRAC and MIPS datasets. We could identify 58 sources out of 87 with a detection in at least one {\it Spitzer} band. Thirty four of them were included in our radial velocity survey which also included 7 stars identified as non-members by \citet{Hunsch03}. 

In the case of NGC~2451 A, sixteen \citet{Hunsch03} members are included in our radial velocity survey, which confirms the membership status of 10. In NGC2451 B we have 13 \citet{Hunsch03} members in our spectroscopic sample. Eight of them are confirmed as members by the radial velocity data while two have radial velocities indicating that they are members of NGC2451 A. We can reject 3  \citet{Hunsch03} members based on their radial velocity data. There is a third group of \citet{Hunsch03} stars that contains objects that might be members of either of the clusters. Six of them are included in our radial velocity survey. Based on the radial velocity data we confirm membership of 5 objects (4 in NGC2451 A and 1 in NGC2451 B). We included the \citet{Hunsch03} members without radial velocities in our final sample if they were detected in at least one of the {\it Spitzer} bands (3 and 10 for NGC2451 A and B, respectively). They are marked with filled triangles in Fig \ref{fig:cmds}. These additions make the total number of cluster members 69 and 131 for NGC2451 A and B, respectively. We detect 22 members at 24$\mu$m, 47 at 8 and 4.5$\mu$m, and 49 at 5.8 and 3.6 $\mu$m in NGC2451 A. In NGC2451 B we detect 20 members at 24$\mu$m, 92 at 8 $\mu$m, 96 at 5.8 $\mu$m, 94 at 4.5 $\mu$, and 98 at 3.6 $\mu$m. Our detection rate in the IRAC bands is about 70\% in both clusters while at 24 $\mu$m it is 32\% in NGC2451 A and 15\% in NGC2451 B, respectively. These detection rates are very simlar to that of \citet{Gorl07} for the $\sim$35 Myr old cluster NGC2457. In the following we use only the sources that are detected in all four IRAC bands or detected at 24$\mu$m. 

\begin{figure*}[!t]
\begin{center}
\includegraphics[scale=0.6]{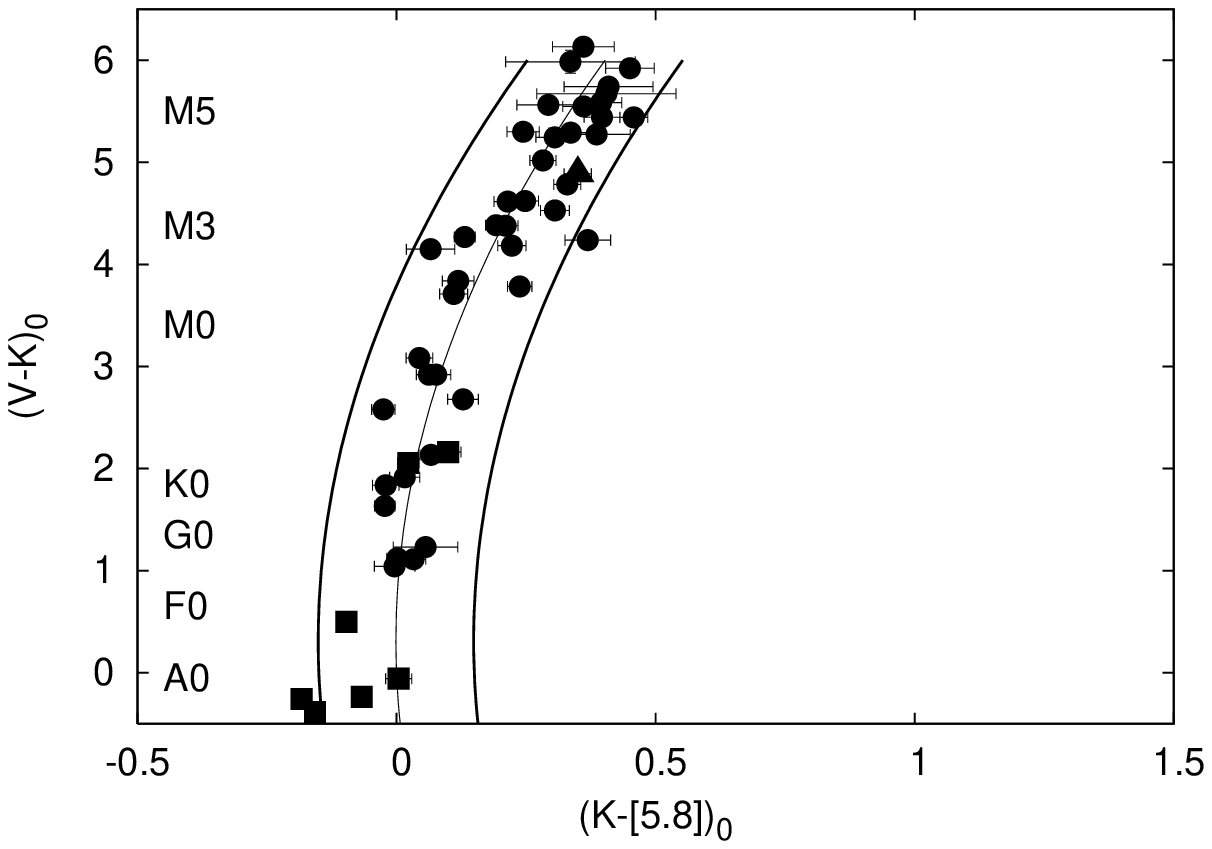}
\includegraphics[scale=0.6]{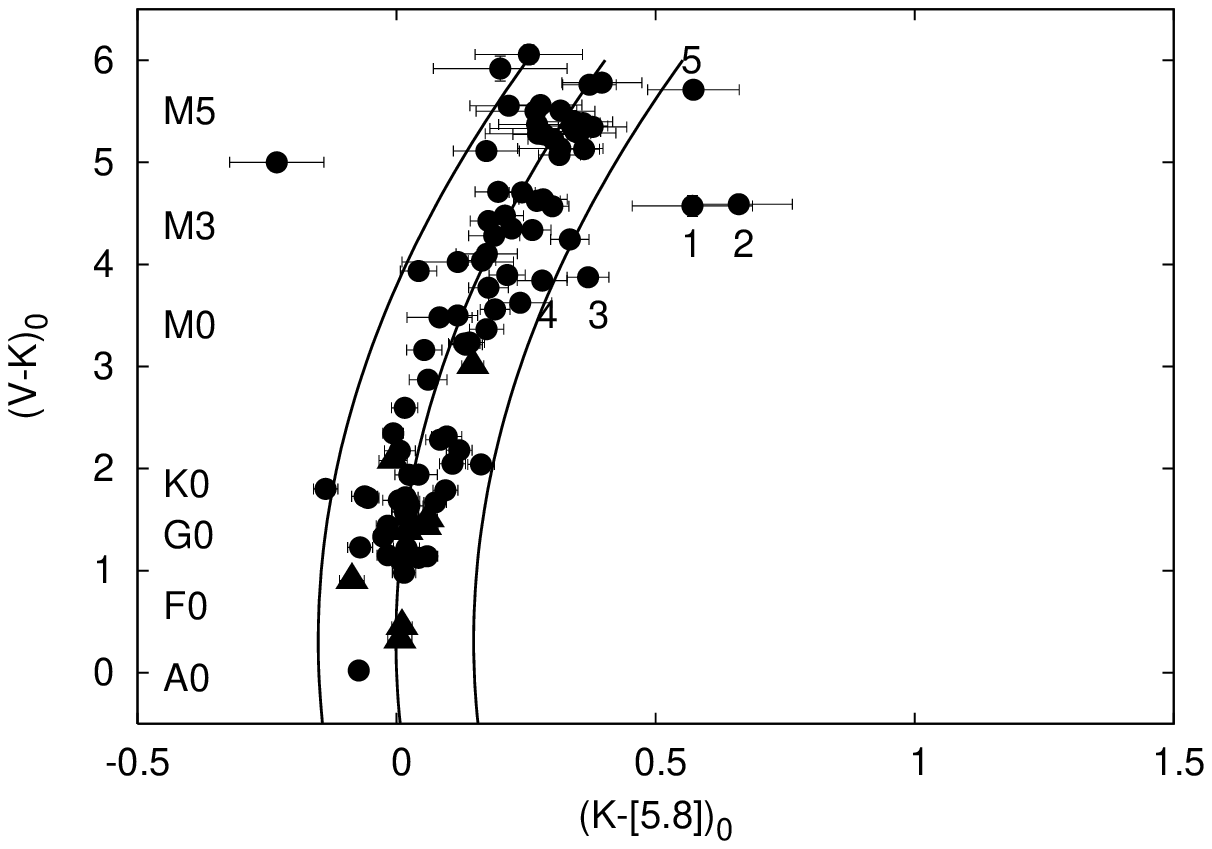}
\caption{The same as Fig \ref{fig:ccds} but for V-K vs K-[5.8].}
\label{fig:ccds2}
\end{center}
\end{figure*}

\section{Identifying excess candidates}

\subsection{IRAC excesses}

The four IRAC bands covering the  3.6$-$8 $\mu$m regime of the spectrum trace dust from the near-sublimation temperature of 1000 to 400 K, corresponding to radii within 1 AU for solar-type stars. We can identify 8 $\mu$m excesses from the IRAC data, if we plot K-[8] vs V-K. Fig \ref{fig:ccds} shows the V-K vs K-[8] color-color diagram of NGC2451 A (left panel) and B (right panel). The large baseline of the V-K color makes this filter combination an ideal choice for photometric spectral type determination given the low reddening of the clusters. There is a monotonic relationship between increasing V-K and stellar spectral type that results in a narrow locus, extending from  early to late type stars; the scatter in this locus increases for V-K$>$5  due to measurement errors. Objects to the right of this locus are stars with excesses at 8 $\mu$m. 

The robust determination of cluster membership allows us to establish a good color-color relation for 50-80 Myr clusters by fitting a 2nd degree polynomial to the locus of the V-K - K-[8] and V-K - K-[5.8] diagrams. We merged the data of the two clusters to increase the number of sources in the fit and required both colors of an A0 star to be 0. Eq. 3 and 4 show the results of the fit which is also shown with a solid line on Fig \ref{fig:ccds} and \ref{fig:ccds2}. The rms of the datapoints around the fit is 0.05 in both colors.

\begin{eqnarray}
(K_S-[8])_0 &=& 0.0115\pm0.0011 (V-K_S)_0^2 \nonumber \\
&&-5.89\times10^{-5}\pm0.0053(V-K_S)_0 \\
(K_S-[5.8])_0&=&0.0125\pm0.0012 (V-K_S)_0^2 \nonumber \\
&&-0.0078\pm0.0060(V-K_S)_0
\end{eqnarray}

\begin{figure*}[!t]
\begin{center}
\includegraphics[scale=0.6]{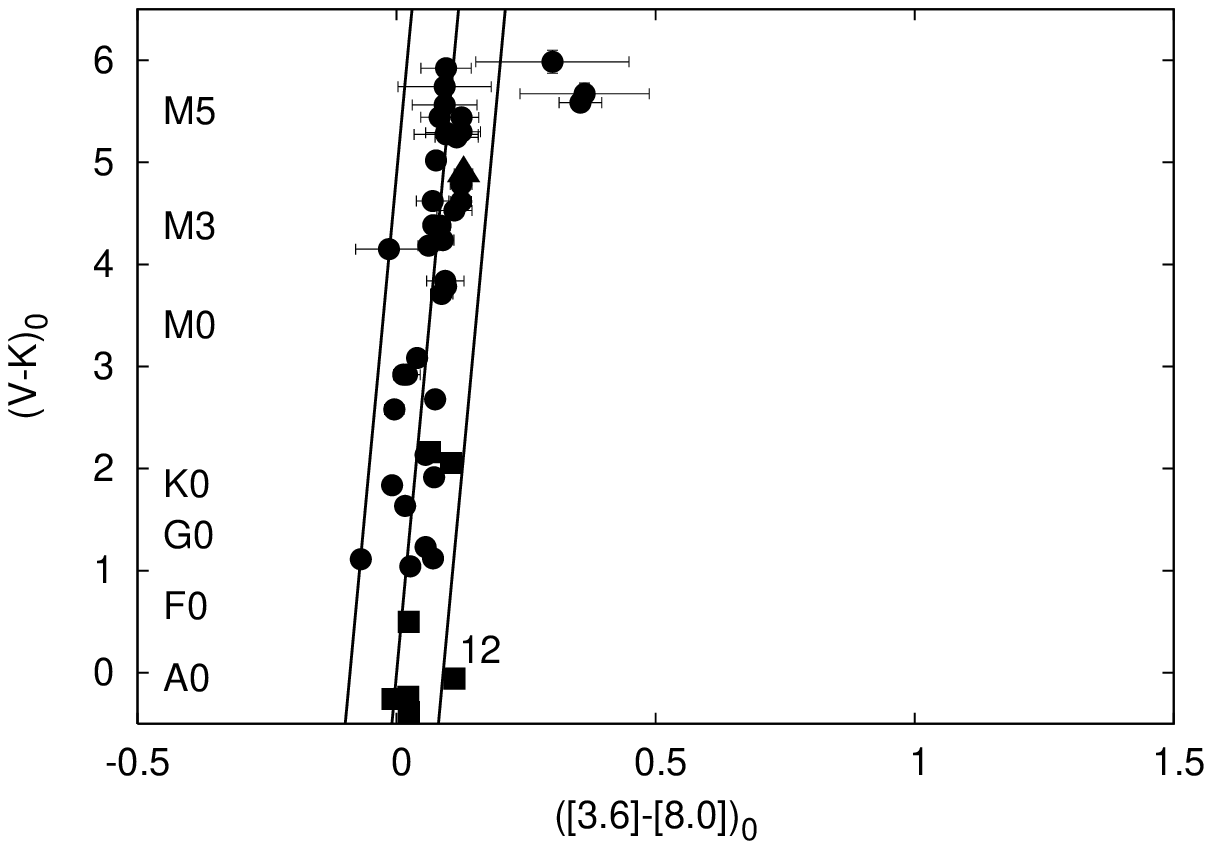}
\includegraphics[scale=0.6]{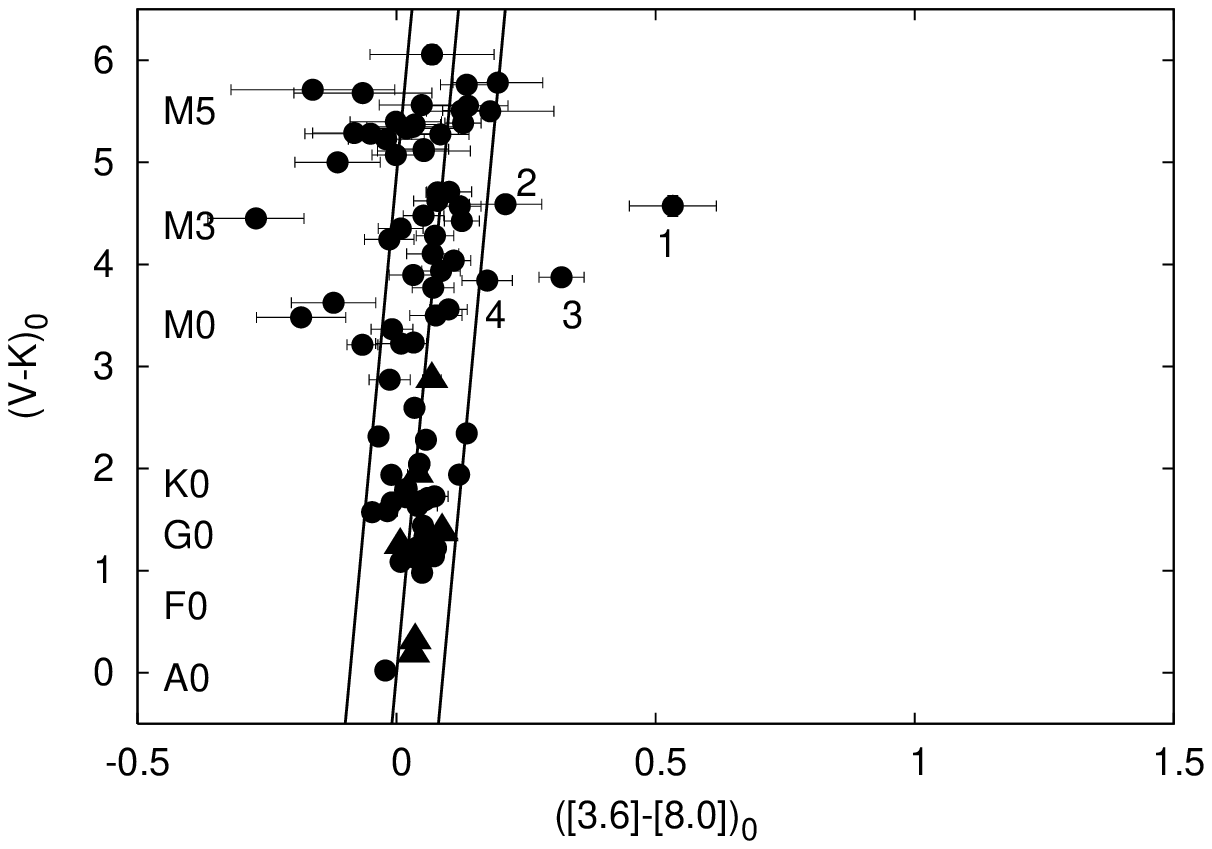}
\caption{The same as Fig \ref{fig:ccds} but for V-K vs [3.6]-[8.0].}
\label{fig:ccds4}
\end{center}
\end{figure*}

\begin{figure}[h]
\begin{center}
\includegraphics[scale=0.35]{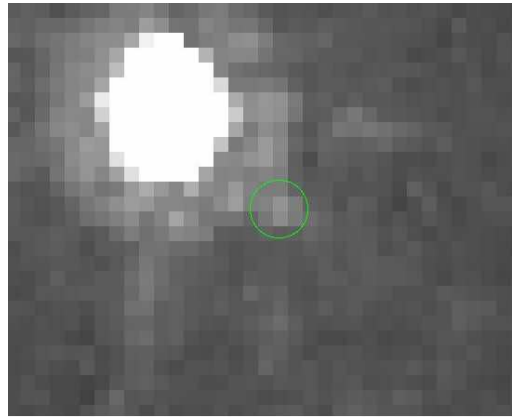}
\caption{Location of star \#2. The green circle shows the aperture size that was used for the photometry}
\label{fig:pair}
\end{center}
\end{figure}

No star with 8 $\mu$m excess (relative to K) was detected in NGC2451 A and only two candidate excess stars (\#1 and \#2) and another two weak possible excess candidates (\#3 and \#4) were detected in NGC 2451 B. We also plot the same diagrams but using the IRAC 5.8 $\mu$m channel on Fig \ref{fig:ccds2}. This diagram allows us to check whether our 8 $\mu$m detections are real or are just a result of some anomaly in the image (e.g. a cosmic ray hit). Both 8 $\mu$m excess candidates are confirmed by having excess at 5.8 $\mu$m also. However star \#2 has a close companion that might compromise the 8 $\mu$m and 5.8 $\mu$m photometry (see Fig \ref{fig:pair}). Stars \#3 and \#4 are also situated slightly to the right of the main locus of sources in V-K vs K-[5.8], however, \#3 has a larger excess at 5.8 $\mu$m so its status as an excess candidate is questionable. None of the four 8 $\mu$m excess candidates were detected at 24$\mu$m.

\begin{figure*}[!t]
\begin{center}
\includegraphics[scale=0.6]{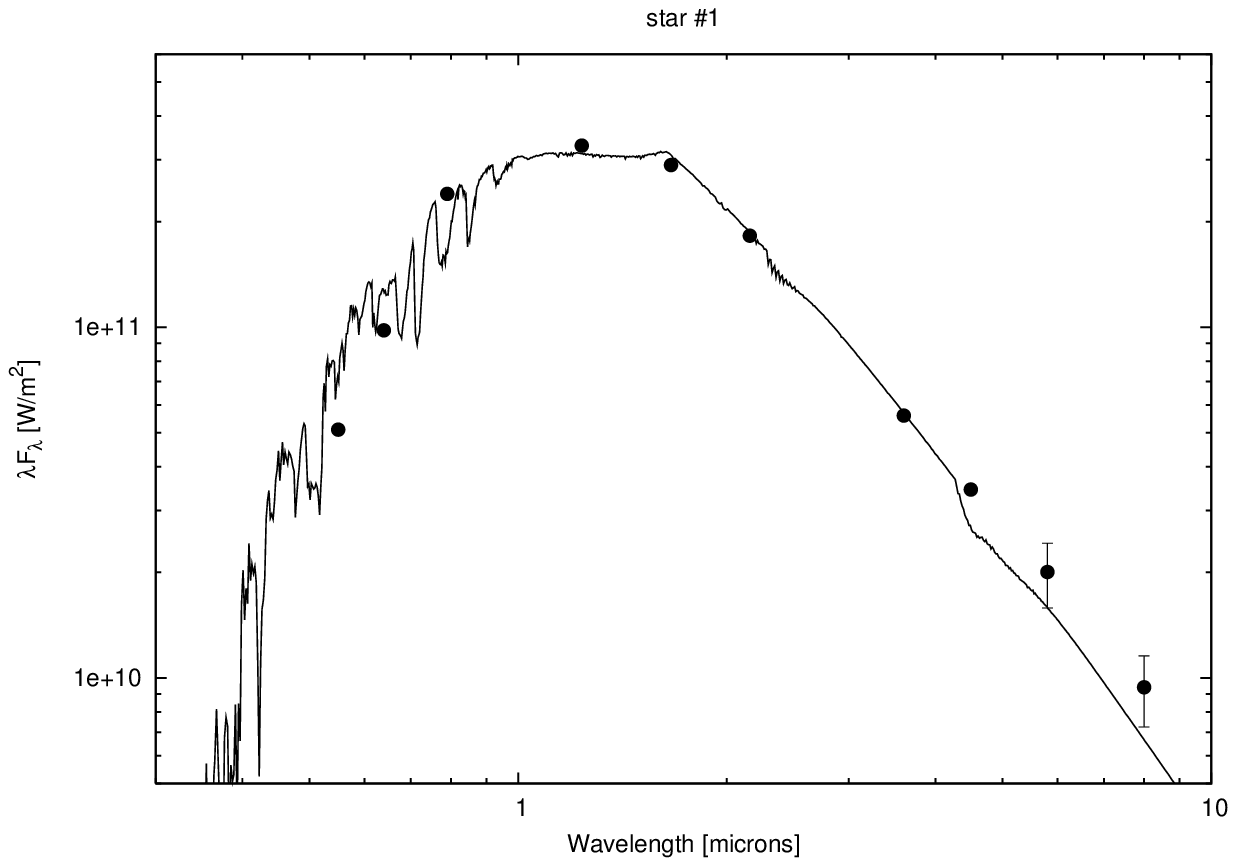}
\includegraphics[scale=0.6]{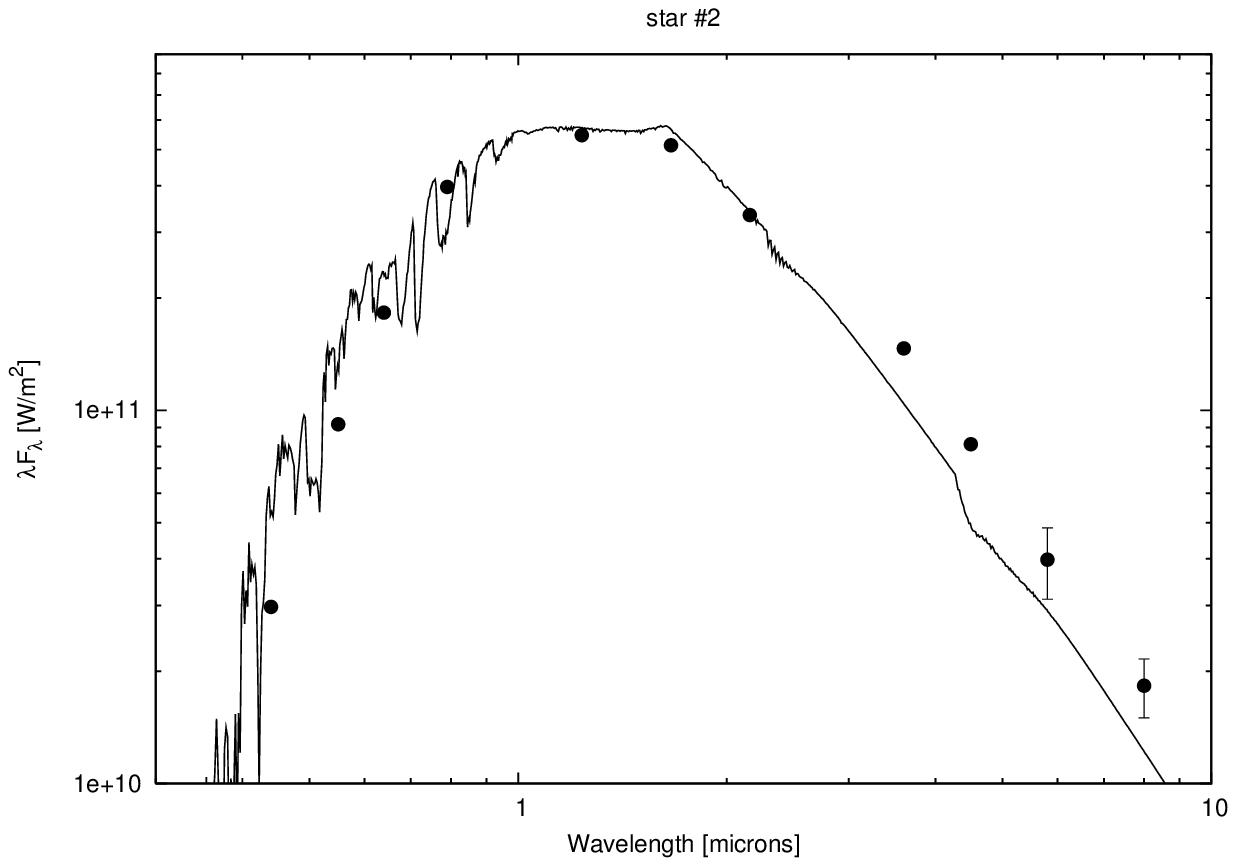}

\includegraphics[scale=0.6]{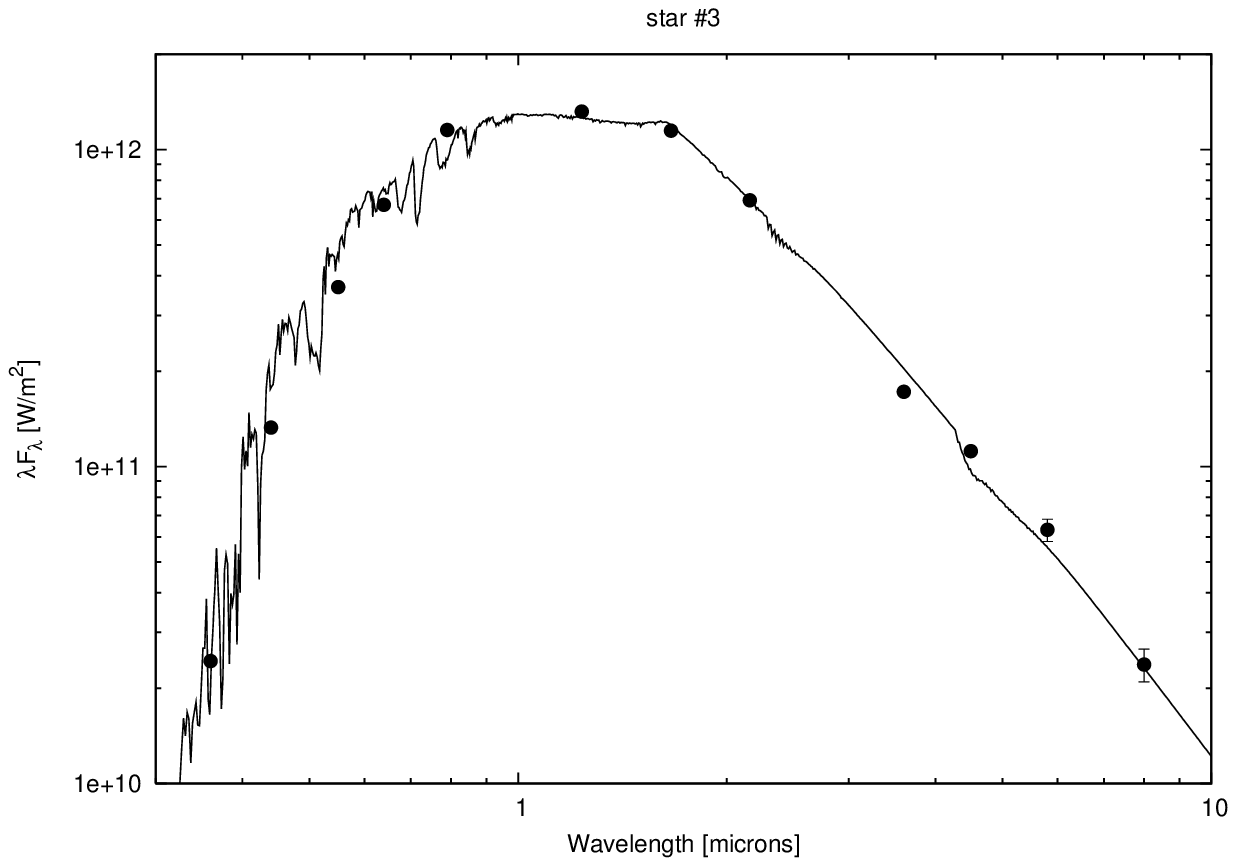}
\includegraphics[scale=0.6]{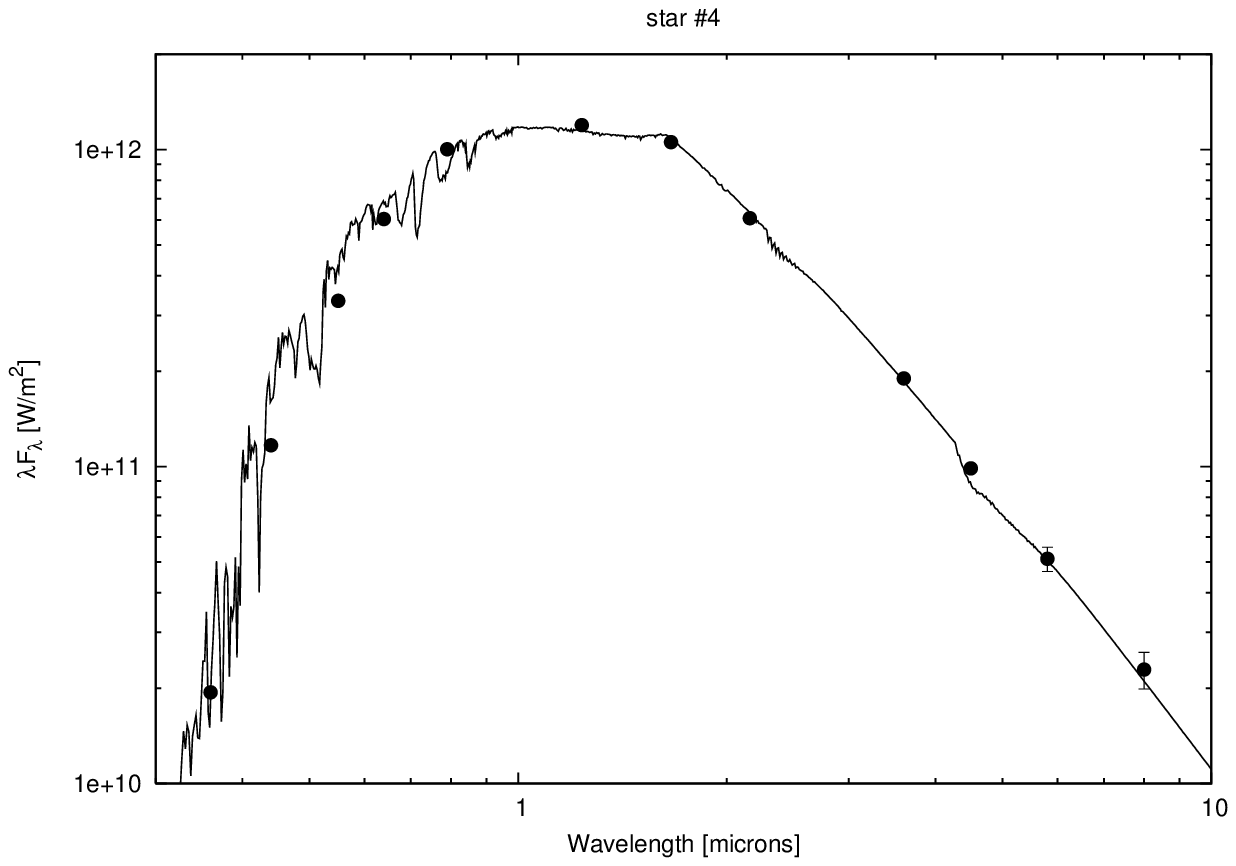}
\caption{The SEDs of the 8$\mu$m excess candidates together with the fitted Kurucz atmospheris models. Errorbars at 5.8 and 8.0 $\mu$m show the 3 $\sigma$ photometric error.}
\label{fig:SEDs1}
\end{center}
\end{figure*}

We fitted a Kurucz model to the short wavelength portion of the sources' SEDs. Fig. \ref{fig:SEDs1} shows the results of the fit. Based on their SEDs we can conclude that only one of the four candidates (\#1) is a true excess source. The SED of \#2 shows that the photometry of this source is contaminated by the nearby bright star because the excess extends over all IRAC bands. The SED of \#3 shows no excess at all and that of \#4 has a marginal one. The significance of the excess for \#4 calculated from the photometric errors and the rms of the fit is slightly lower than 3$\sigma$. 

We noticed that there is a small hint of an 8 $\mu$m excess on the SED of the 24$\mu$m excess star \#12 (see next section) and all but one of the A (or earlier) type stars have K-[8] colors less than zero although they are inside the 3$\sigma$ limit of the trendline in Fig \ref{fig:ccds}. To check the origin of this slight discrepancy we replaced K with [3.6] and used the V-K vs [3.6]-[8.0] diagram to detect excesses (Fig. \ref{fig:ccds4}). A linear relation between the V-K and the [3.6]-[8.0] colors can be found for the photospheric sources

\begin{equation}
([3.6]-[8])_0= 0.01842\pm0.001277(V-K_S)_0
\end{equation}

\noindent with rms scatter of 3\%. We also show this relation along with the 3$\sigma$ limits in Fig. \ref{fig:ccds4}.

In this diagram there is no systematic deviation from the main relation and it seems that \#12 has a marginal excess at 8 $\mu$m relative to 3.6$\mu$m. The significance of the excess is about 4$\sigma$. There are some excess candidates among the late M type stars of NGC2451 A but their significance is less than 3 $\sigma$. In the case of NGC2451 B (Fig. \ref{fig:ccds4} right panel) the diagram clearly shows that the only viable 8 $\mu$m excess candidate is \#1. Although \#3 shows some [3.6]-[8.0] excess, the SED of the source does not support it.

\begin{figure*}[t]
\begin{center}
\includegraphics[scale=0.6]{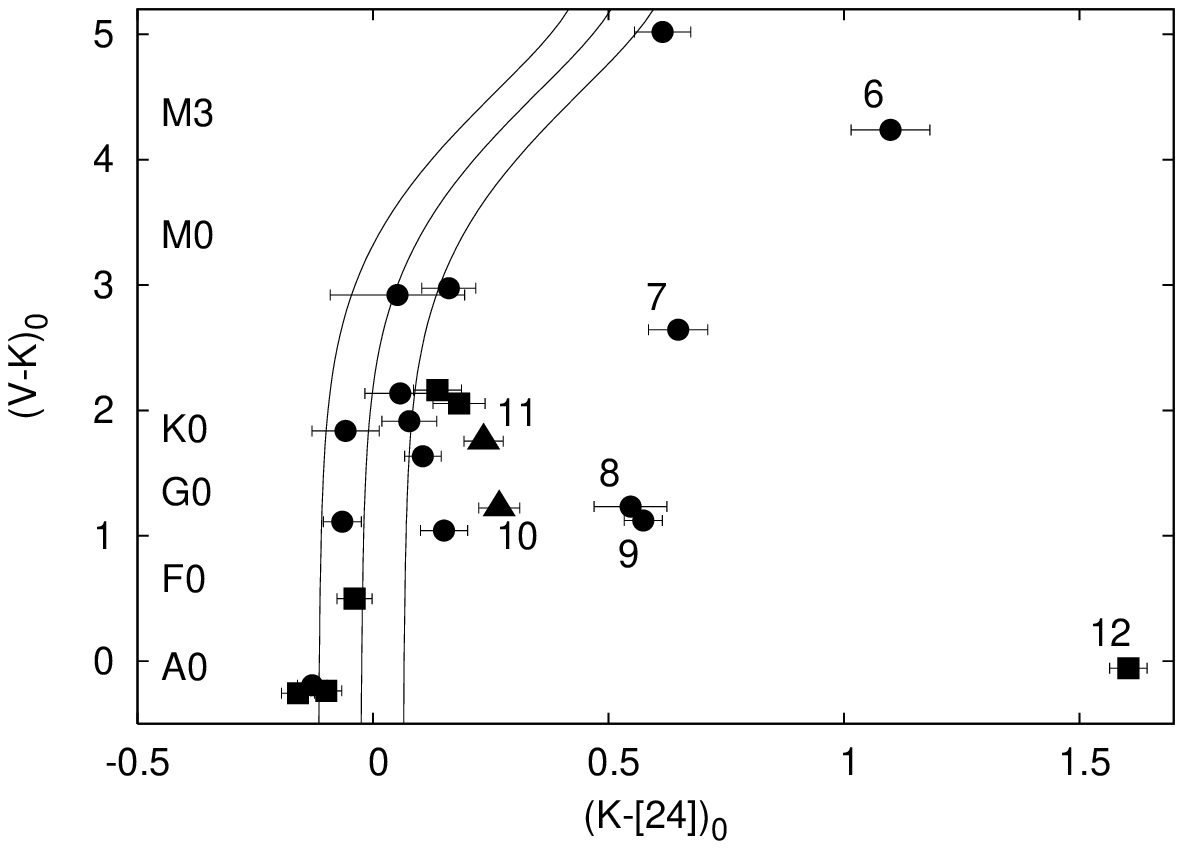}
\includegraphics[scale=0.6]{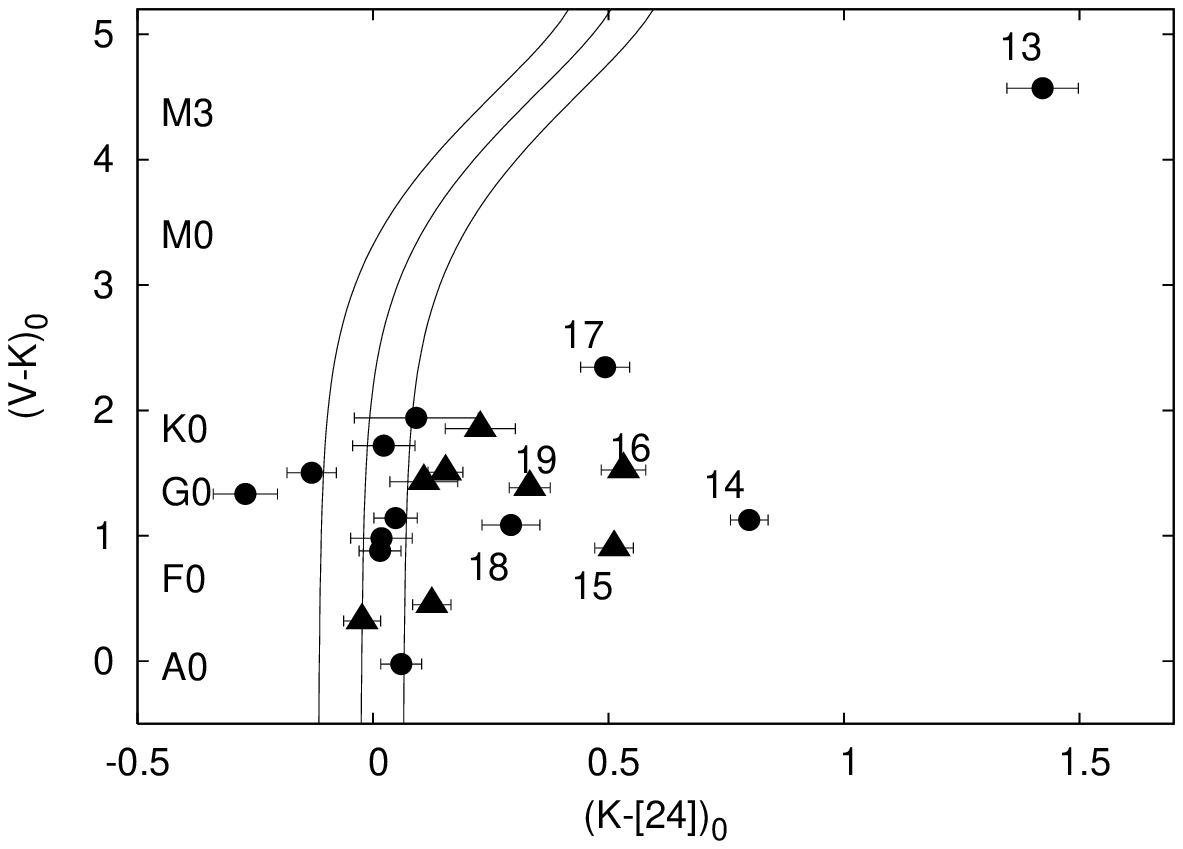}
\caption{The same as Fig \ref{fig:ccds} but for V-K vs K-[24]. The solid lines represent the trendline (see text) and its $3 \sigma$ limits}
\label{fig:ccds3}
\end{center}
\end{figure*}
To summarize, at 8 $\mu$m we have less than 2\% excess fraction in NGC2451 A (one marginal excess) and a less than 1\% excess fraction in NGC2451 B (one clean excess). This result is very similar to that of \citet{Gorl07} suggesting that IR excess shortward of 24 $\mu$m is very rare in clusters older than 20-30 Myr and indicates that the disk within 10 AU is cleared of dust for nearly all stellar systems older than 30 Myr.     

\subsection{MIPS excesses}

\begin{figure*}[!t]
\begin{center}
\includegraphics[scale=0.6]{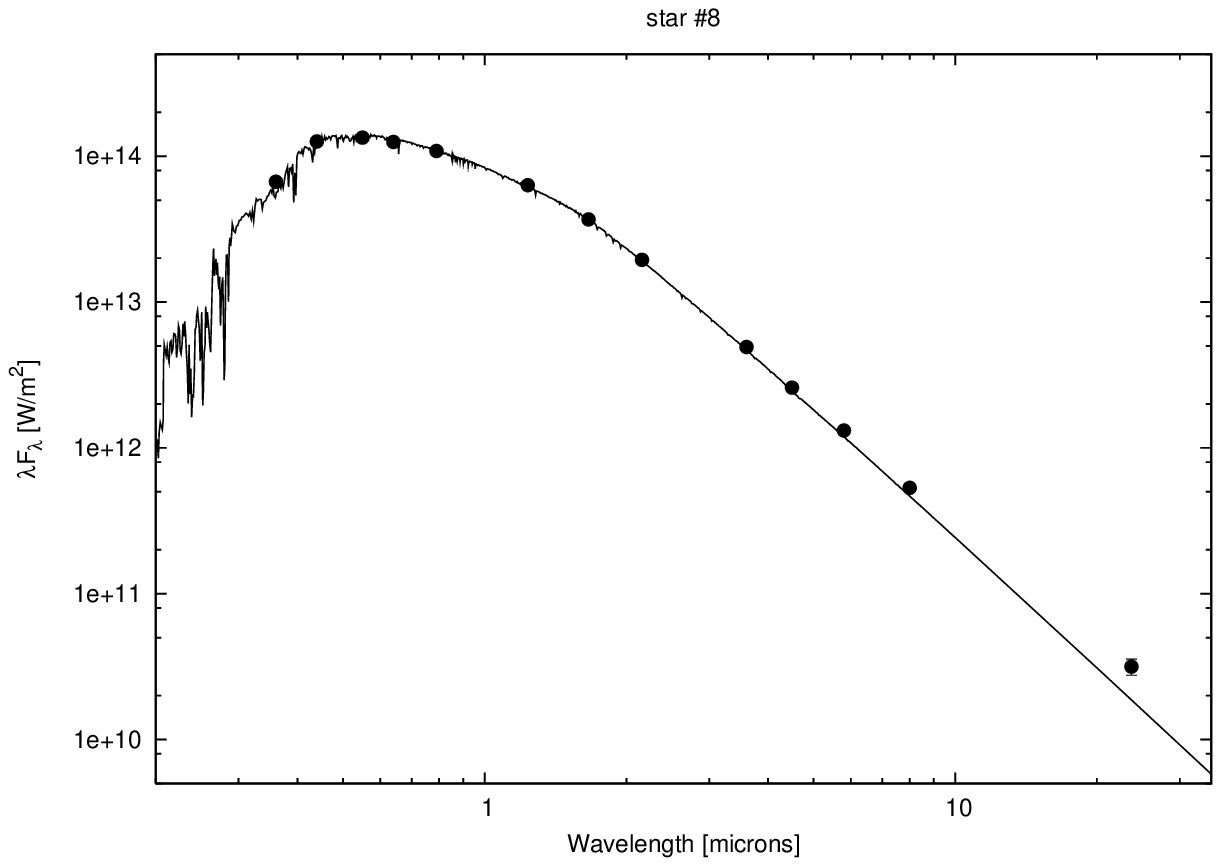}
\includegraphics[scale=0.6]{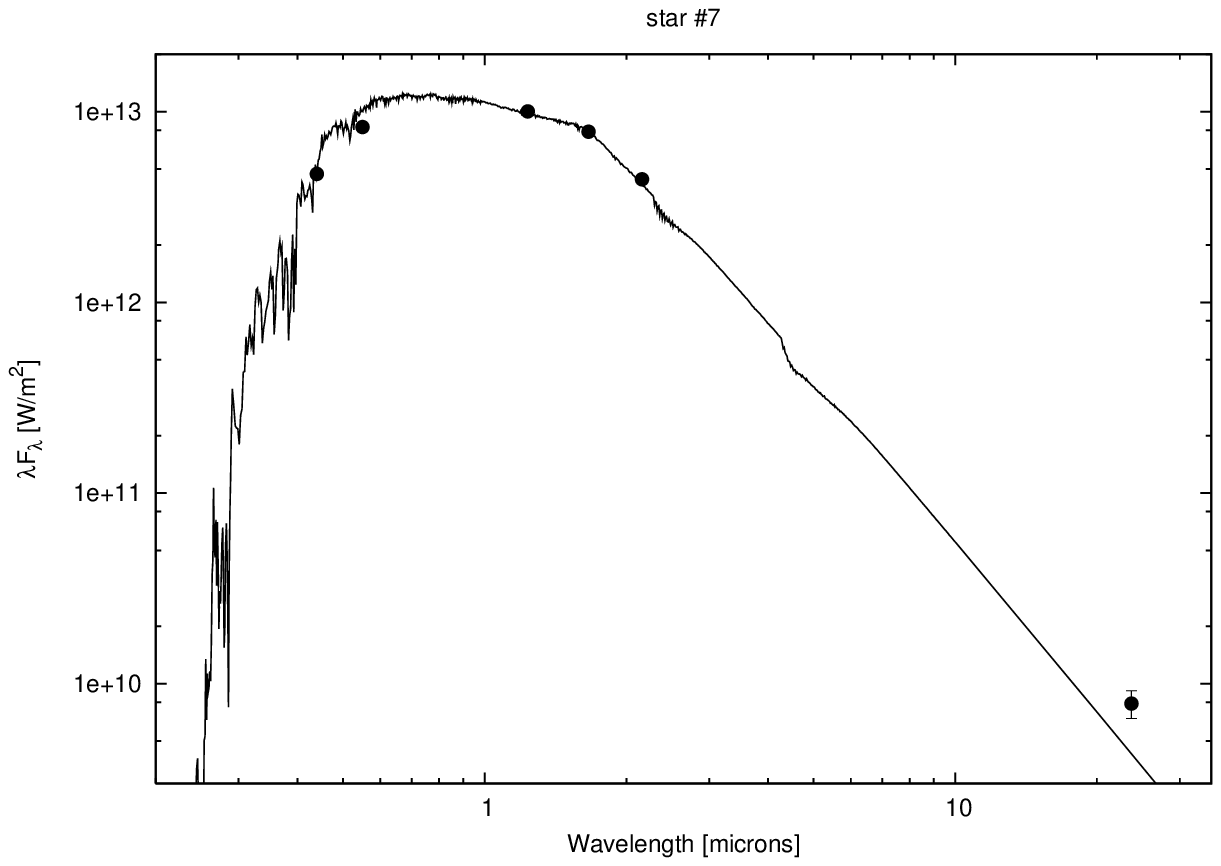}

\includegraphics[scale=0.6]{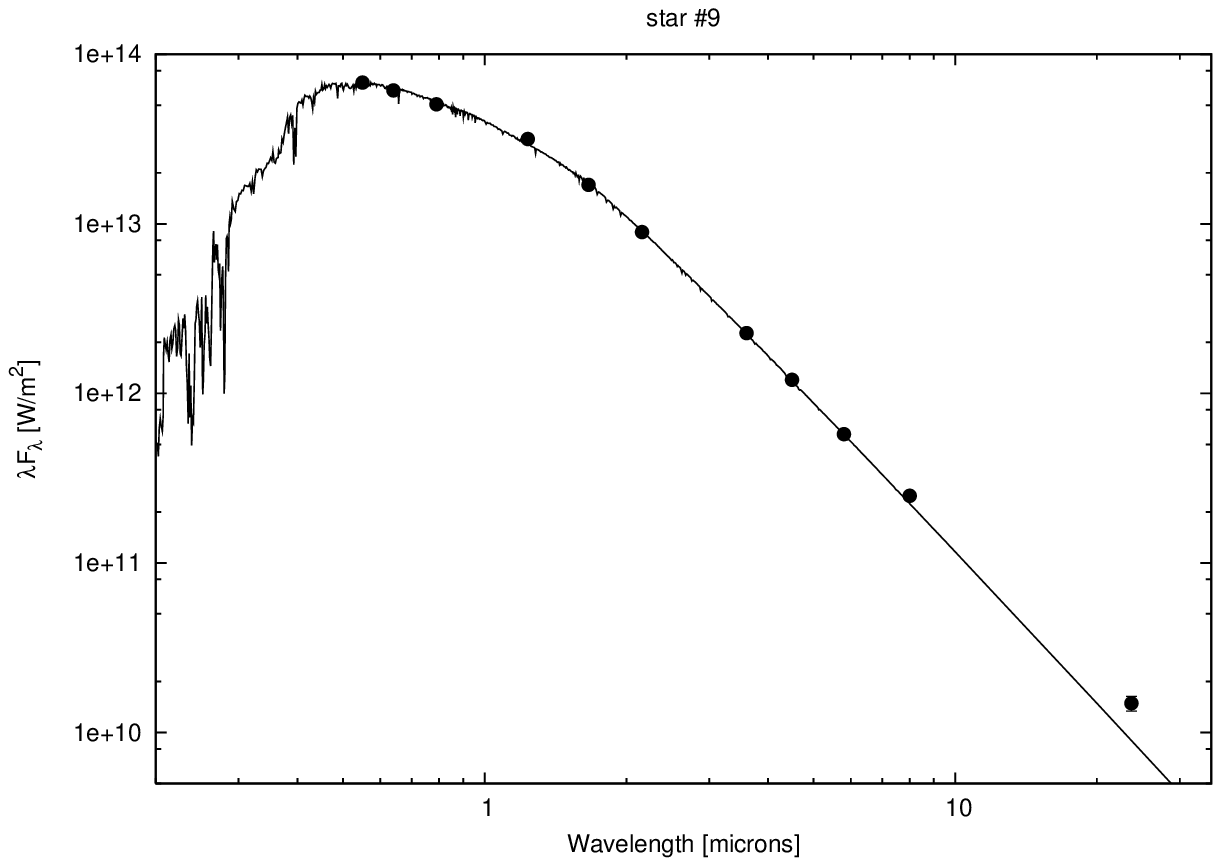}
\includegraphics[scale=0.6]{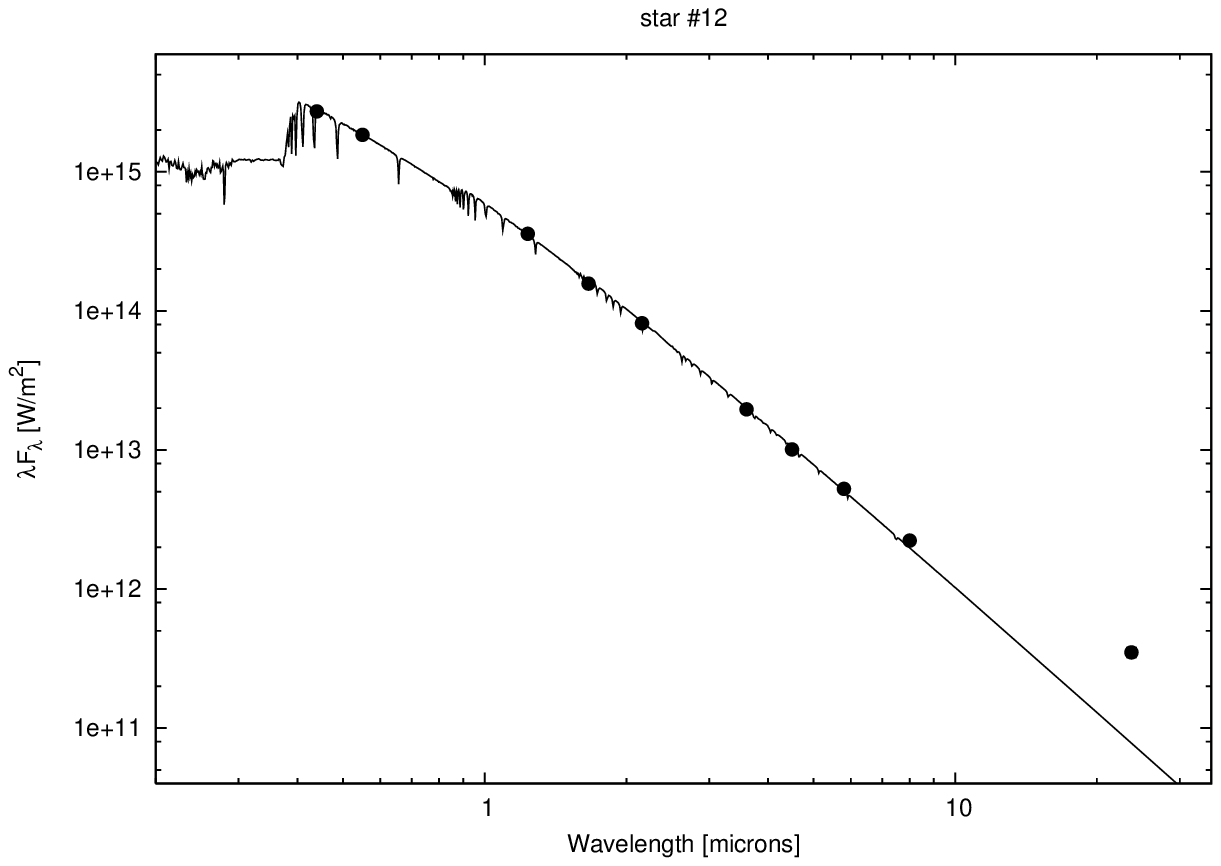}

\includegraphics[scale=0.6]{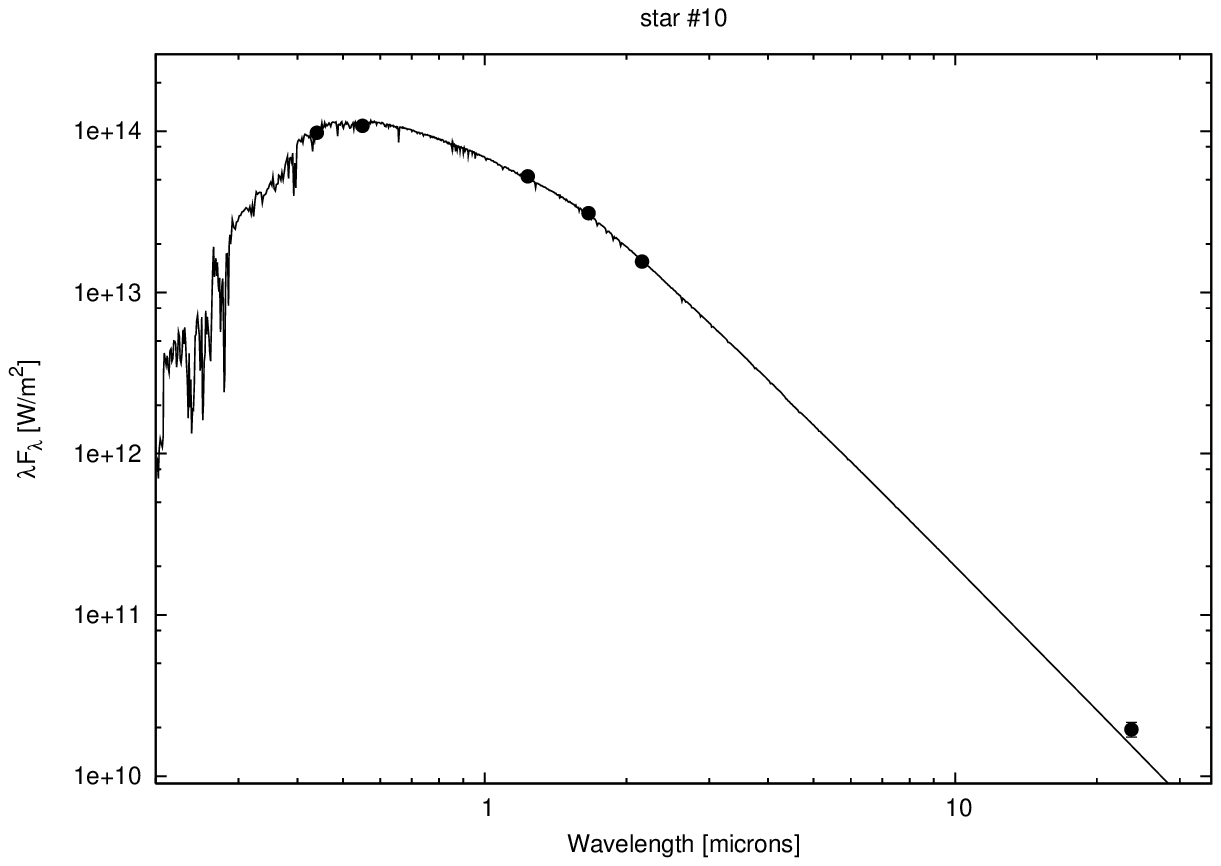}
\includegraphics[scale=0.6]{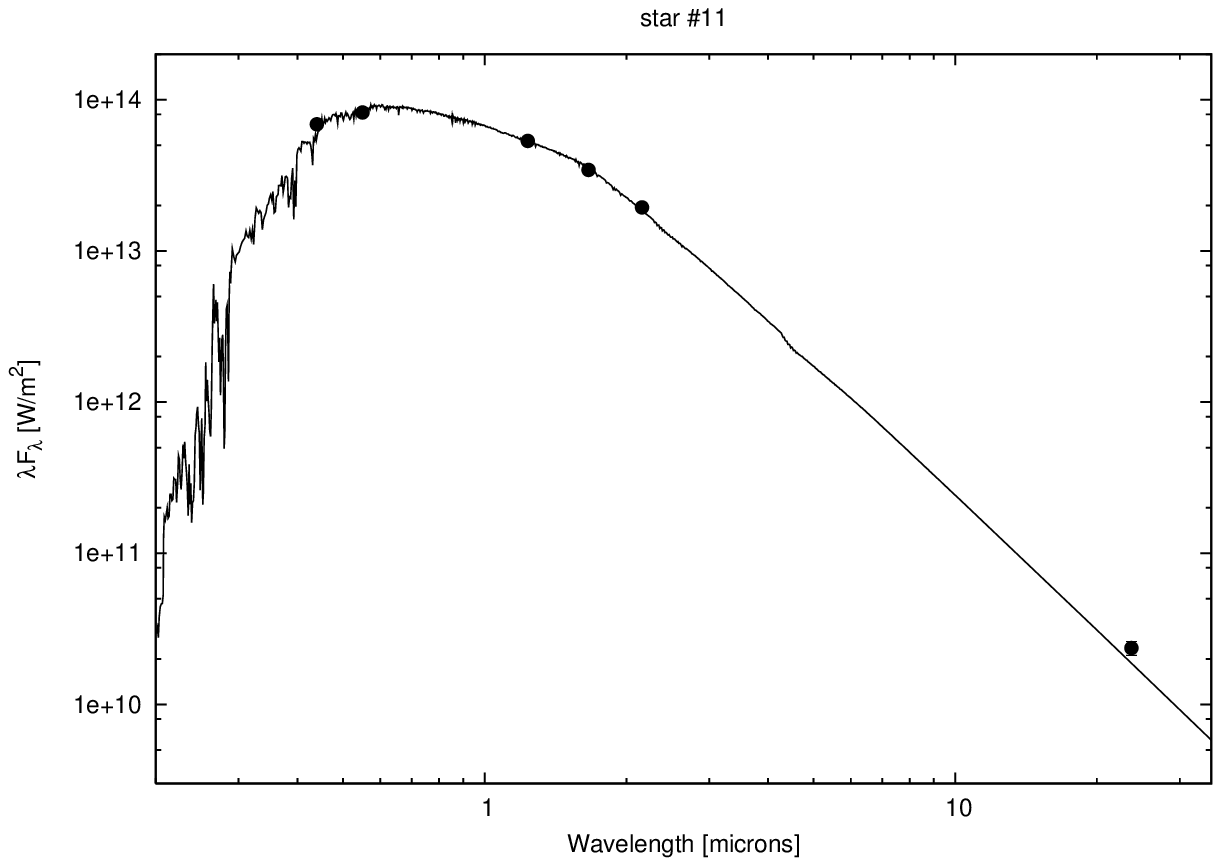}

\includegraphics[scale=0.6]{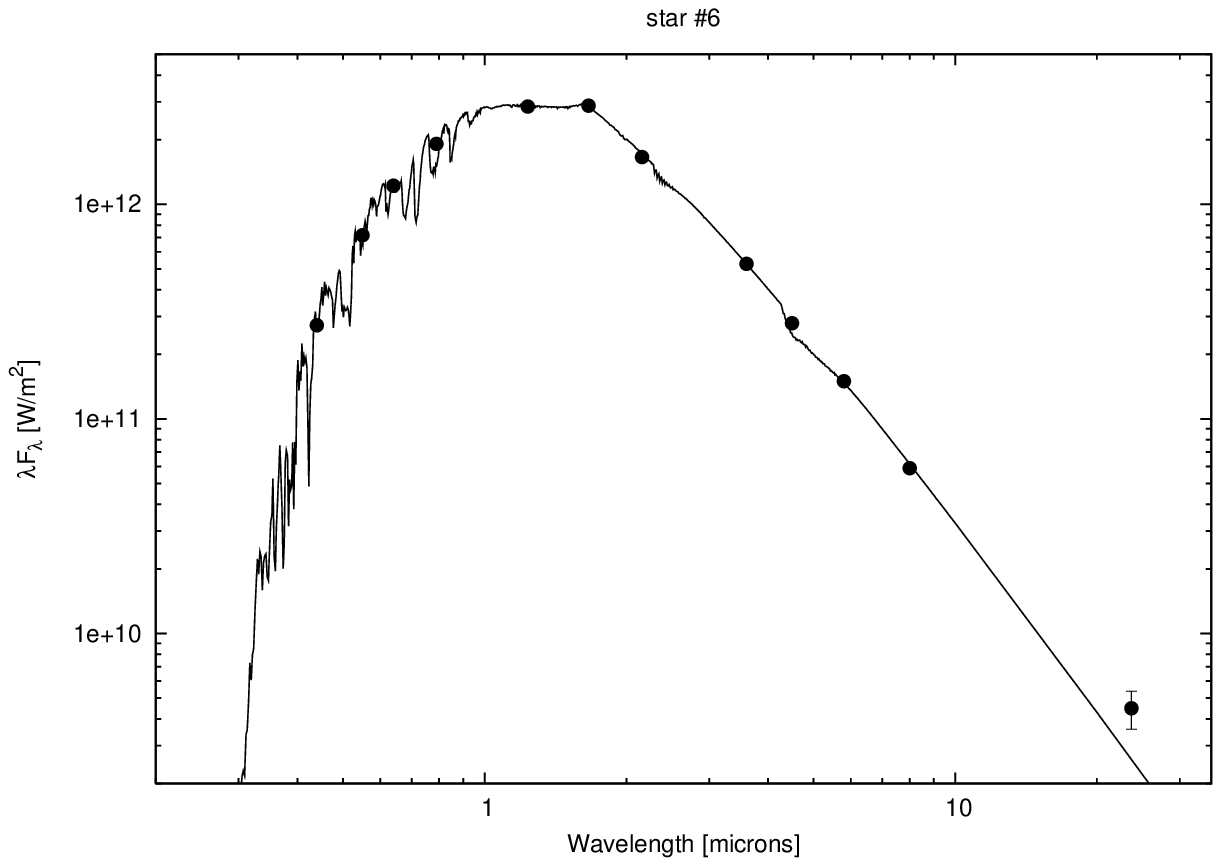}
\caption{The SEDs of the 24$\mu$m excess candidates in NGC2451 A together with the fitted Kurucz atmospheric model. The errorbars represent the 3 $\sigma$ photometric error.}
\label{fig:SEDs2}
\end{center}
\end{figure*}

\begin{figure*}[!t]
\begin{center}
\includegraphics[scale=0.6]{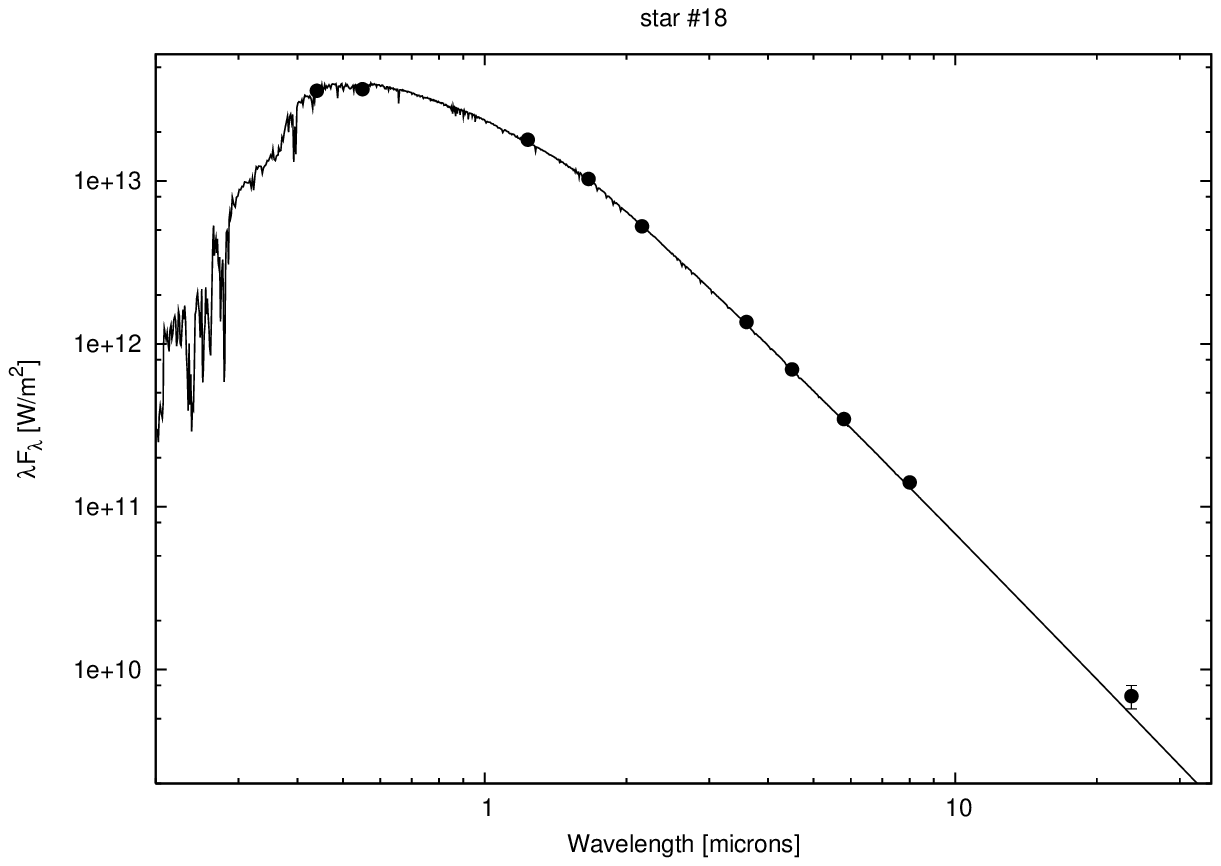}
\includegraphics[scale=0.6]{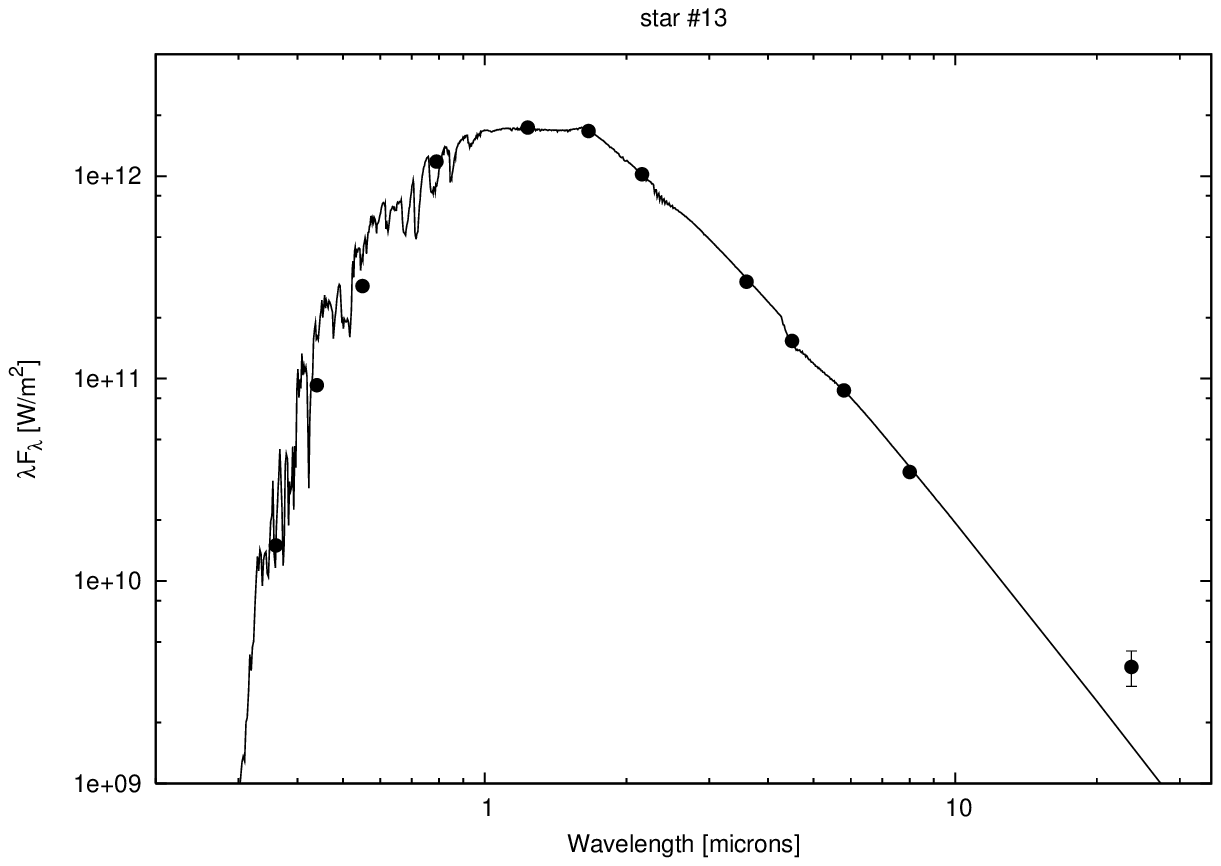}

\includegraphics[scale=0.6]{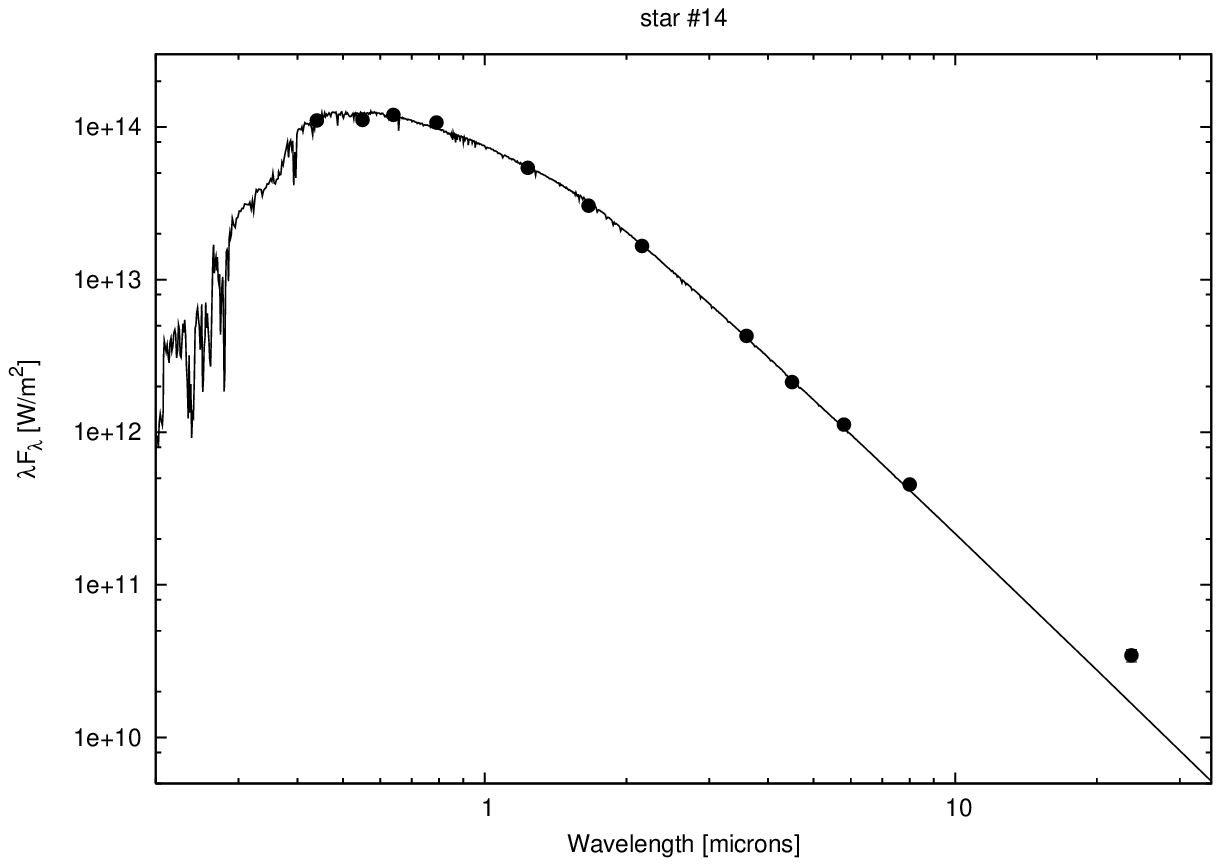}
\includegraphics[scale=0.6]{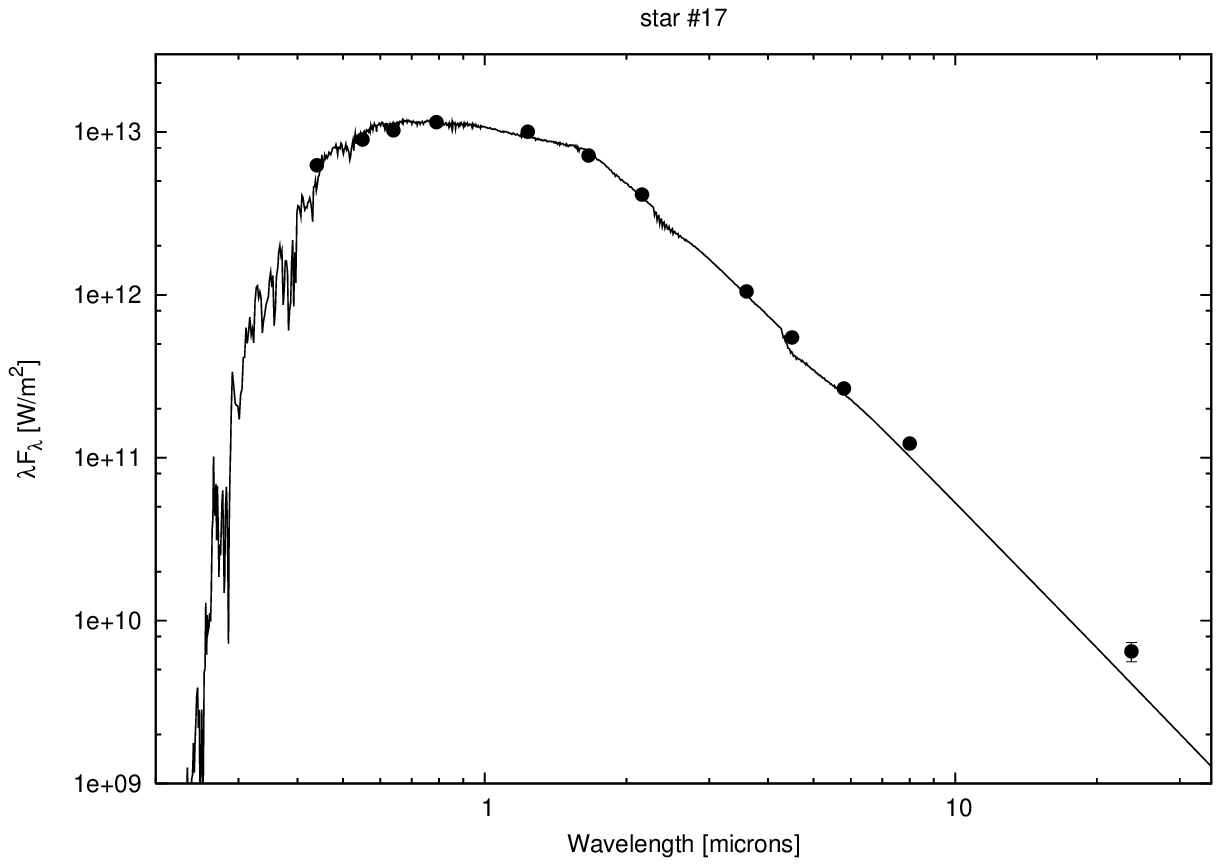}

\includegraphics[scale=0.6]{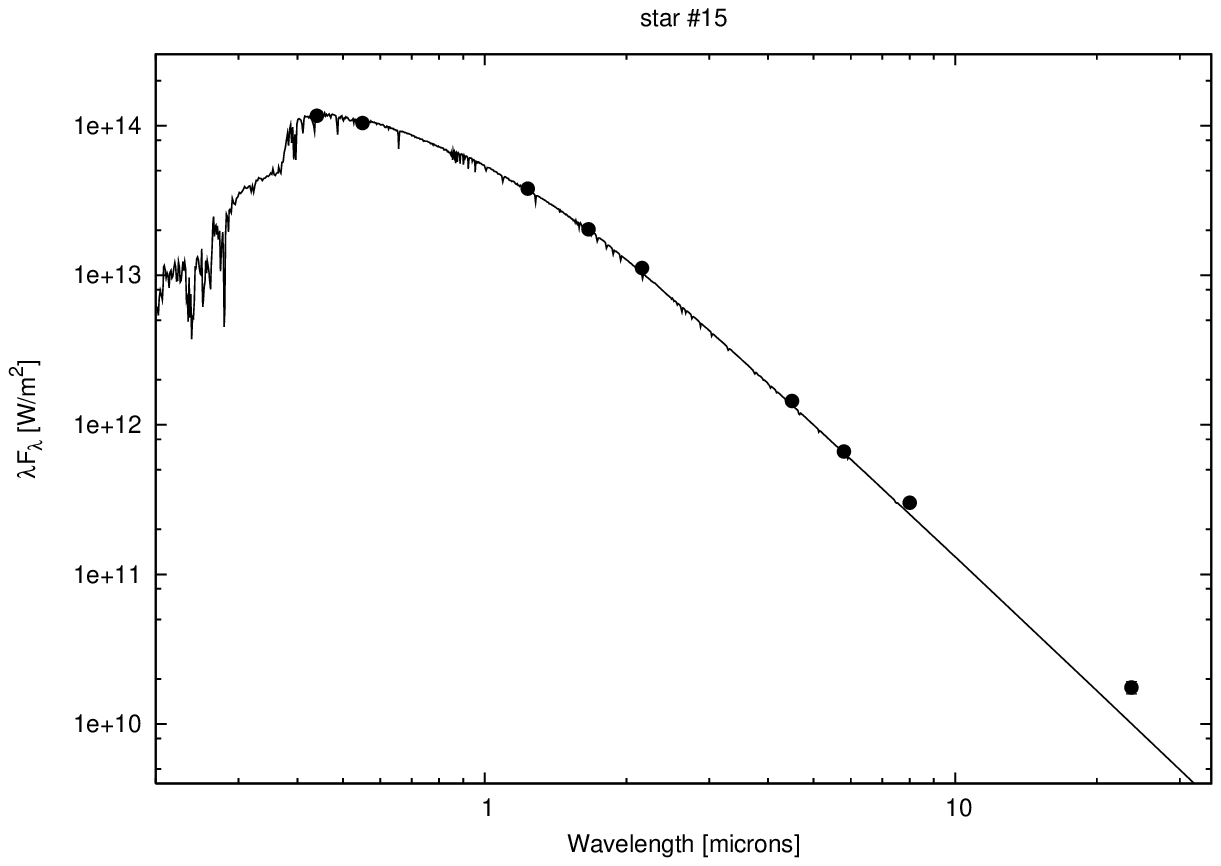}
\includegraphics[scale=0.6]{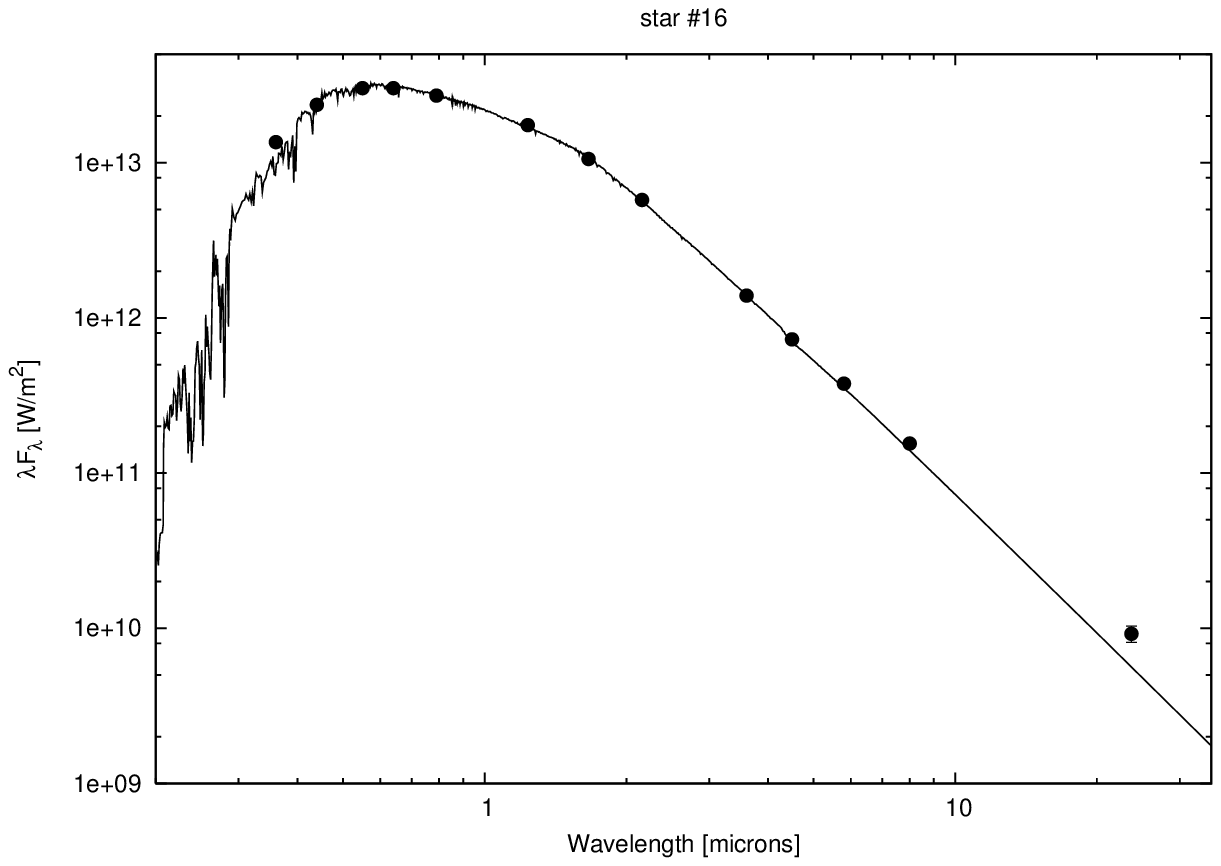}

\includegraphics[scale=0.6]{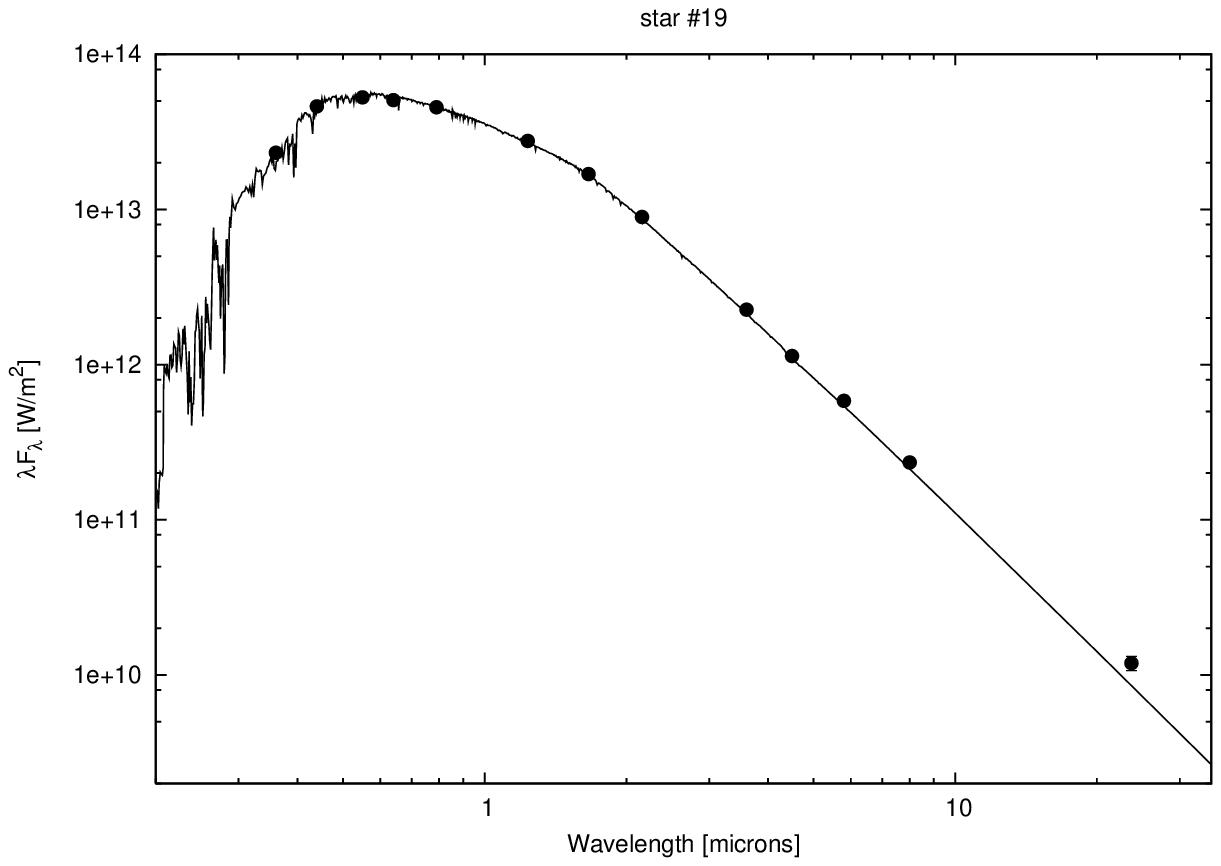}
\caption{The SEDs of the 24$\mu$m excess candidates in NGC2451 B together with the fitted Kurucz atmospheric model.  The errorbars represent the 3 $\sigma$ photometric error.}
\label{fig:SEDs3}
\end{center}
\end{figure*}

\setcounter{table}{6}
\begin{deluxetable*}{ccccccc}
\tablecolumns{7}
\tablewidth{0pc}
\tablecaption{Measured and predicted photospheric (K-[24])$_0$ colors of cluster members with 24 $\mu$m excesses\label{tab:24exces}}
\tablehead{
\colhead{ID}& \colhead{RA}&\colhead{DEC}&\colhead{(K-[24])$_0$}&\colhead{$\sigma$((K-[24])$_0$})&\colhead{(K-[24])$_0^{phot}$}&\colhead{significance of}\\
\colhead{}&\colhead{J2000}&\colhead{J2000}&\colhead{mag}&\colhead{mag}&\colhead{mag}&\colhead{excess}
}
\startdata
\cutinhead{NGC2451 A}
\#6&116.04942 & -37.73994 & 1.099 & 0.084 & 0.271& 9$\sigma$\\
\#7&116.74747 & -38.37820 & 0.648 & 0.063 & 0.023& 9$\sigma$\\
\#8&116.31223 & -38.19624 & 0.547 & 0.077 & -0.018& 7$\sigma$\\
\#9&116.23017 & -38.07956 & 0.574 & 0.041 & -0.019& 12$\sigma$\\ 
\#10&116.53718 & -37.39910 & 0.268 & 0.043 & -0.018& 5$\sigma$\\
\#11&116.10025 & -38.59977 & 0.235 & 0.042 & -0.011& 5$\sigma$\\
\#12&116.18701 & -38.05376 & 1.604 & 0.040 &  -0.024& 33$\sigma$\\
\cutinhead{NGC2451 B}
\#13&116.67488 & -37.98390 & 1.422& 0.076 & 0.353& 13$\sigma$\\ 
\#14&115.95562 & -37.89574 & 0.799 & 0.040 & -0.022& 17$\sigma$\\
\#15&116.21267 & -37.66869 & 0.496 & 0.041 &  -0.023& 11$\sigma$\\
\#16&116.50776 & -38.00752 & 0.516 & 0.047 & -0.022& 10$\sigma$\\
\#17&116.44107 & -37.89720 & 0.493 & 0.052 & -0.010& 9$\sigma$\\
\#18&116.74249 & -37.63916 & 0.293 & 0.061 & -0.024& 5$\sigma$\\ 
\#19&116.36051 & -38.19643 & 0.317 & 0.043 & -0.025& 7$\sigma$
\enddata
\end{deluxetable*}

Fig \ref{fig:ccds3} shows the dereddened V-K vs K-[24] for NGC2451 A (left panel) and NGC2451 B (right panel). Because there are too few detections to determine a zero excess locus purely from these data, for this purpose we have used a database of more than 1500 stars derived from nearly all of the Spitzer debris disk programs. The sample has good coverage from -0.4 $<$ V-K $<$ 3.5, with still a number of members up to V-K = 8. A simple polynomial fit has been made to the photospheric color-color locus. The rms scatter around this fit (clipping the positive outliers that have excesses) is only 3\%. The locus itself is therefore very well determined and we can take a typical uncertainty for the measurement of any given star to be 3\% relative to this locus (plus any statistical errors). This locus with 3-$\sigma$ errors (9\%) is plotted in Fig \ref{fig:ccds3}.

It is obvious that excess at 24 $\mu$m is far more common than at shorter wavelengths. We find 9 stars in the case of NGC2451A and 10 stars in the case of NGC2451 B to the right of the photospheric locus. However, only 7 of them in both diagrams (the numbered sources in Fig \ref{fig:ccds3}) are more than 3$\sigma$ distance from the locus (the distance of the negative outlier left of the locus). The measured (K-[24])$_0$ color along with the predicted photospheric K-[24] color of these objects and the significance of the excesses are presented in Table \ref{tab:24exces}. For the prediction of the photospheric colors we used the photospheric locus shown in Fig \ref{fig:ccds3}.  The significance was calculated using the photometric erors and the rms around this locus. To check the validity of our excess candidates (as in the previous section) we fit Kurucz model atmospheres to the short wavelength part of their SEDs. The fits are shown in Fig \ref{fig:SEDs2} and \ref{fig:SEDs3}. In all cases the 24 $\mu$m flux is above the photosphere by more than 3 $\sigma$, confirming the excess.  The fraction of stars with excess at 24 $\mu$m is about 33$\pm 10$ \% (7/22) for NGC2451 A and 36$\pm 10$ \% (7/20) for NGC2451 B. We calculated the disk fractions and errors using Bayesian statistics as described in \citet{Gasp08a}. All but one of our 24 $\mu$m excess candidates have spectral type F or later (based on their V-K colors and fitted SEDs).  We detect only one star earlier than F0 with 24 $\mu$m excess in NGC2451 A and none in NGC2451 B. 

Our excess fractions among solar type stars are 35$\pm12$ \% (5/15) in NGC251 A and 39$\pm 12$ \% (6/16) for NGC 2451 B, and for early type stars are 29$\pm17$\% (1/5) and 20$\pm 20$ \% (0/3) for NGC2451 A and B, respectively.

\section{Discussion}
\subsection{Excess fractions}

\begin{figure*}[!t]
\begin{center}
\includegraphics[scale=0.6]{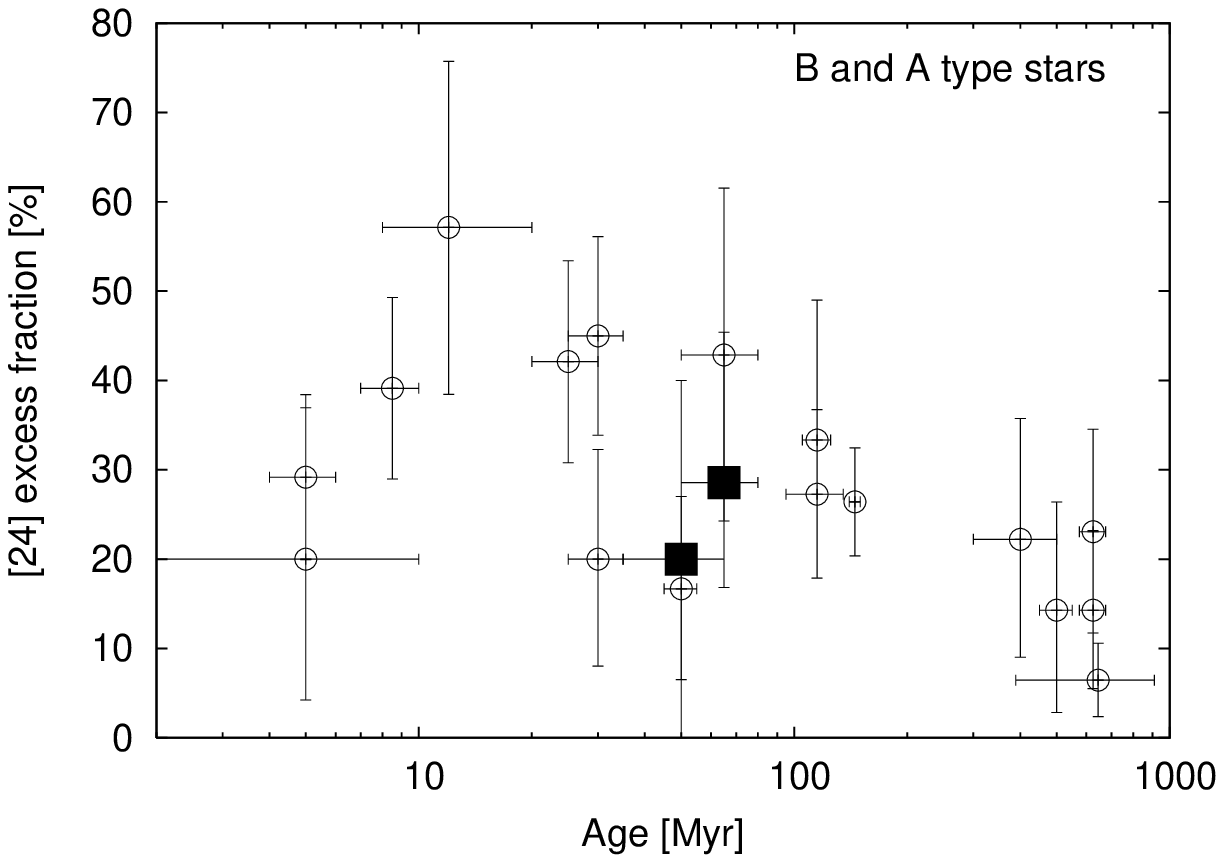}
\includegraphics[scale=0.6]{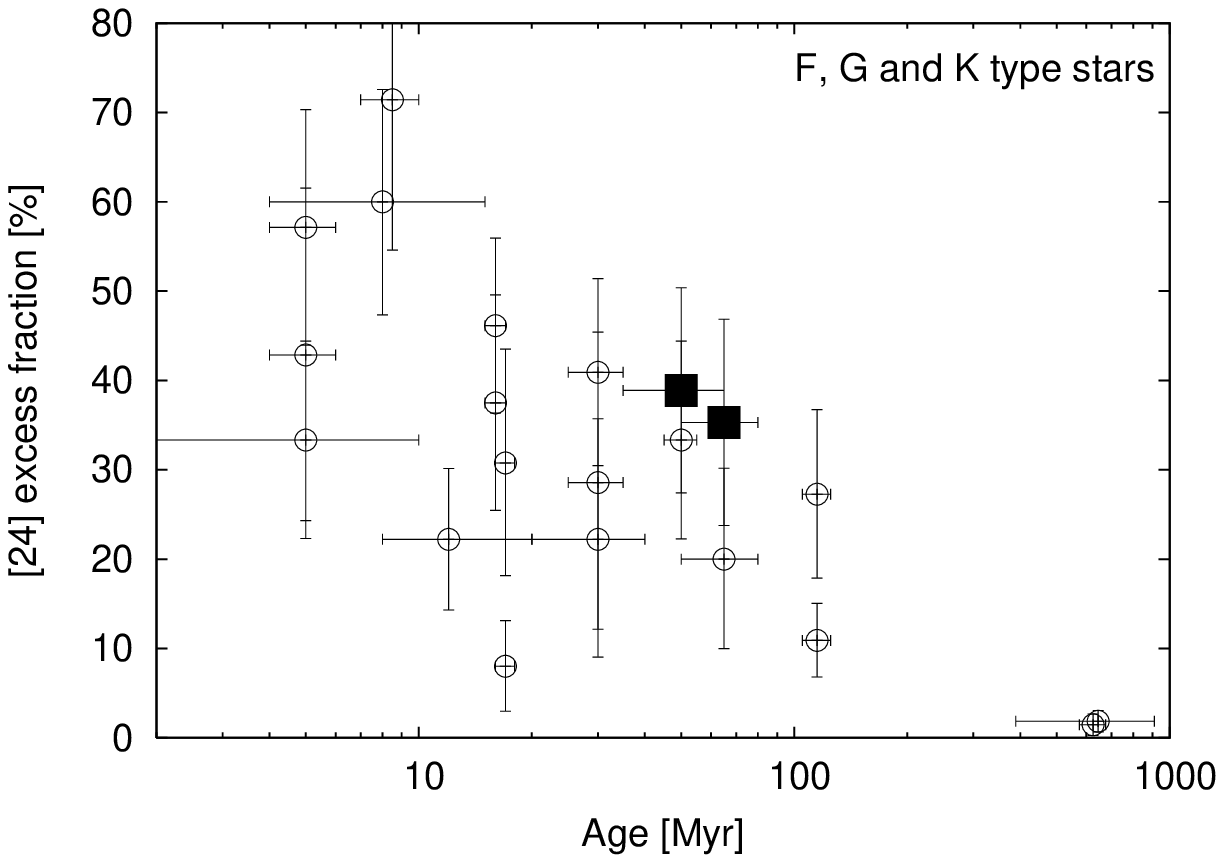}
\caption{The fraction of star with 24$\mu$m excess vs age in open clusters. Left panel: early type stars. Right panel: solar type stars. Open circles: data from \citep{Gasp08a}; filled squares: NGC2451 A and B}
\label{fig:disk_freq}
\end{center}
\end{figure*}

\begin{figure*}[t]
\begin{center}
\includegraphics[scale=0.6]{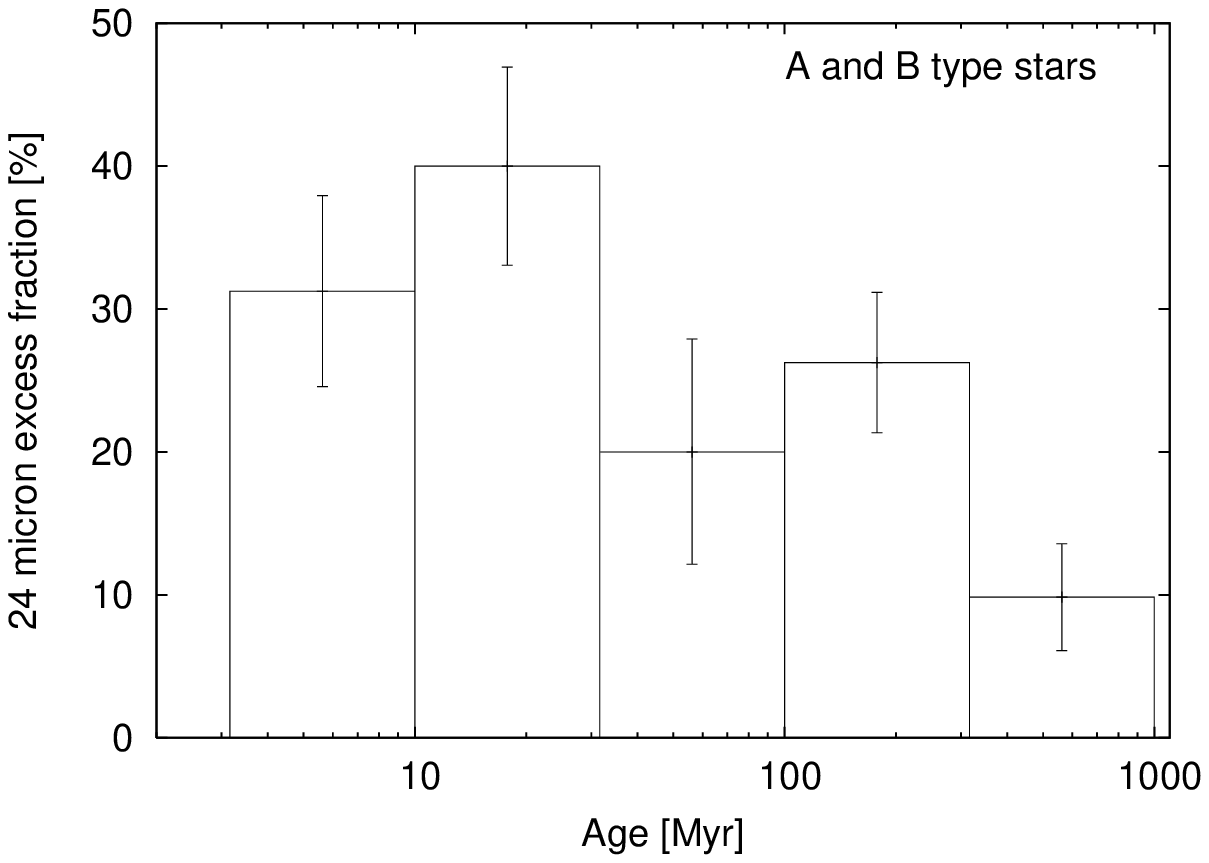}
\includegraphics[scale=0.6]{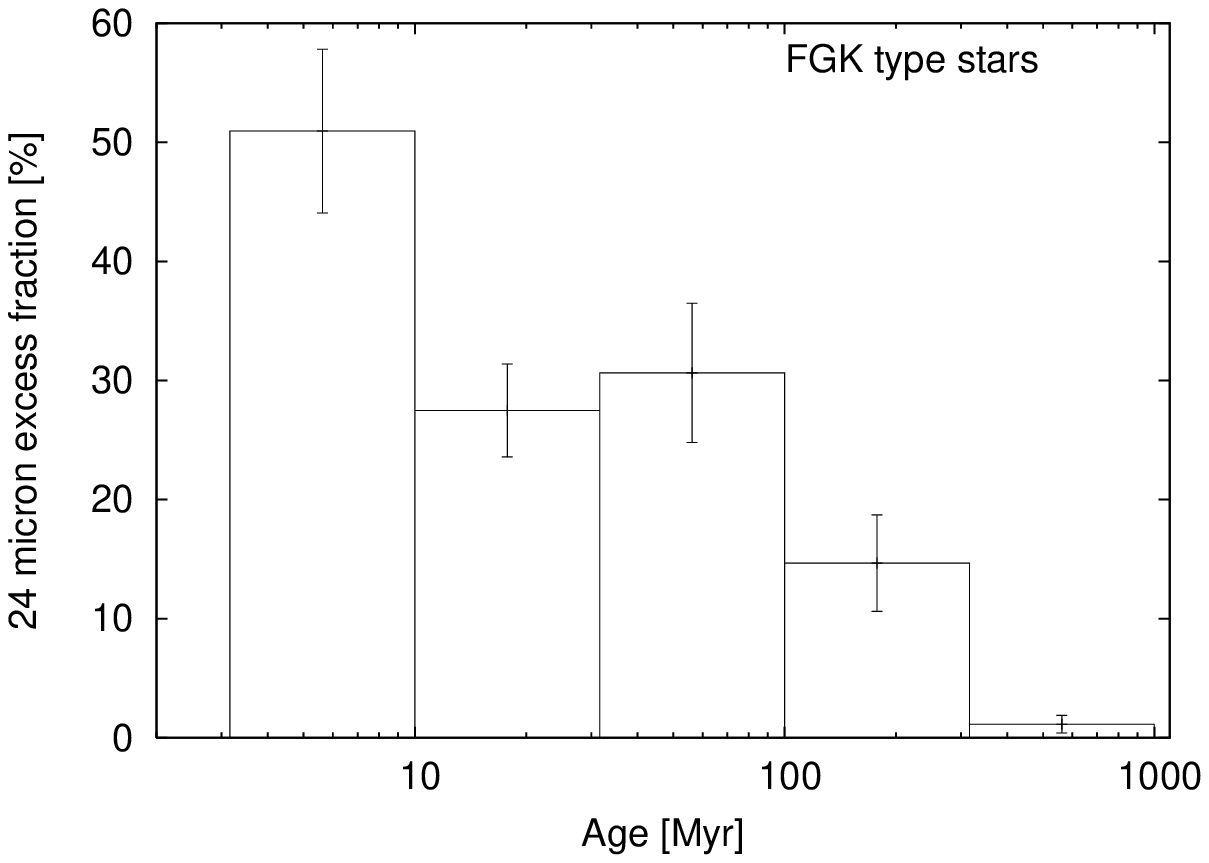}
\caption{The binned excess fraction of open cluster vs age. Left panel: early type stars. Right panel: solar type stars.}
\label{fig:bins}
\end{center}
\end{figure*}

Our results for excess fraction in the solar type stars of NGC2451 B are roughly consistent with other clusters with similar ages such as IC 2391 (50 Myr - 31\% \citep{Sieg07}) or NGC 2547 (35 Myr - 40\%; \citep{Gorl07}). In the case of NGC2451 A our excess fraction is significantly larger than that of M47 (80 Myr - 6\% \citep{Gorl04}) which might indicate that the age of NGC2451 A is closer to 50 Myr rather than 80 Myr. However, the study of M47 has a higher threshold for detection of excesses, which might also account for the lower fraction.

For the early type stars our results are in agreement with those of \citet{Sieg07} and \citet{Su06} who found a very low excess fraction ( about 10\% and 14\%  for IC2391 (50 Myr) and IC2602 (30 Myr), respectively).

Fig \ref{fig:disk_freq} shows the disk fractions as a function of age (similar to Fig 9. of \citet{Gasp08a}) (left panel: early type stars; right panel: late type stars). Open circles show the data from Table 4 and 6 of \citet{Gasp08a} while filled squares designate the two new datapoints from our study. We note that for stars younger than 10 Myr the excess not necessarily indicates the presence of debris but primordial dust left over from the star formation process as pointed out by \citet{Rhee07}, so the first three points in the figure might overestimate the debris disk fraction. An interesting feature of the figures is that there is a dip in the fraction of debris disks in the age range of 30-80 Myr. Outside this range the decay seems to be monotonic. Our two clusters slightly deviate from the main trend in the late type stars also but in a different direction. They are both situated above the main locus of stars along with some other clusters of similar age. Incompleteness might account for the discrepancy. However, even if we remove the faintest stars (\#17 in Fig \ref{fig:ccds3}) in NGC2451 B and the three faintest ones (\#7 and the two photospheric sources around V-K$\sim$3 Fig \ref{fig:ccds3} left panel) in NGC2451 A, the excess fractions (35$\pm 12$ \% and 36$\pm 13$\% for NGC2451 A and B, respectively ) are still slightly above the main locus. 

Unfortunately low number statisics might reduce the reliability of these results. To check whether they are statistically significant we binned the cluster data of \citet{Gasp08a} supplemented with our two clusters into five equal logarithmic age bins 3-10Myr, 10-31.6Myr, 31.6-100Myr, 100-316Myr, 316-1000 Myr.  We show the result of the binning in Fig \ref{fig:bins} (left panel: early type stars;right panel: late type stars).  There is a hint of a drop (about 1 $\sigma$) in the excess fraction of early type stars in the 31.6-100 Myr range then, a rise and another drop after 316 Myr. For the late type stars a small bump breaks the continous decay with about the same significance as the dip in the early type sample. More detailed and very accurate membership study of the clusters in this age range is necessary to decide whether this behavior can be attributed solely to statistical fluctuation.

\subsection{Stars with large excess}

Among all stars in NGC 2451 A/B showing excess at 24$\mu$m, \#12 (HD 62938) has by far the largest; K-[24] = 1.6 corresponding to an excess ratio (R) more than 4 meaning that the 24$\mu$m flux density is more the 4 times larger than the expected photospheric flux level. The MIPS data place a 2$\sigma$ upper limit of 63 mJy on the flux density at 70 $\mu$m, requiring a color temperature between 24 and 70 $\mu$m $\geq$ 95 K. To put this star into context with other results we performed a literature search for main-sequence stars older than 20 Myr with similarly large or larger excess. We found 9 stars with excess ratios larger than 4 at 24$\mu$m. We also included two famous somewhat younger large excess stars HD39060 ($\beta$ Pic) and HD181296 ($\eta$ Tel) for comparison. They are listed in Table \ref{tabl:largeexc}. The excess of HD21362, a B6 star, is due to free-free emission so we rejected this star from our final sample.

\begin{figure}[b]
\begin{center}
\includegraphics[scale=0.6]{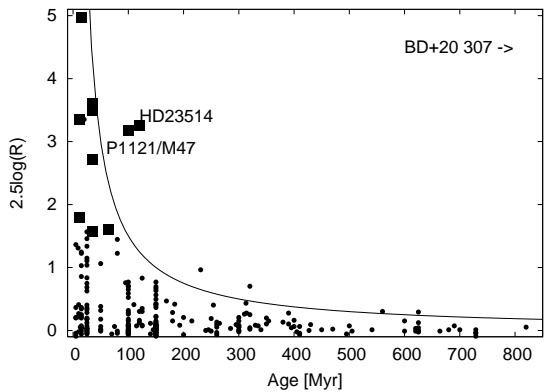}
\caption{24$\mu$m excess vs age. Squares: stars with large excess from Table \ref{tabl:largeexc}, dots: stars from \citet{Riek05}, solid line: the normal decay trend from infrared excesses }
\label{fig:largeexc}
\end{center}
\end{figure}

\begin{deluxetable*}{lcccc}
\tablecolumns{5}
\tablewidth{0pc}
\tablecaption{Stars with large 24$\mu$ excess\label{tabl:largeexc}}
\tablehead{
\colhead{Name}&\colhead{SpT}&\colhead{Age}&\colhead{excess ref.}&\colhead{age ref}
}
\startdata
HD21362$^{\dagger}$ & B6Vn & 80 Myr & \citet{Riek05}& \citet{Song01}\\
HD39060 & A5V & 12 Myr & \citet{Riek05}&\citet{Orteg02,Orteg04}\\
HD109573 & A0V & 10-20 Myr & \citet{Riek05}&\citet{Zuck04a}\\
NSV17775$^{\ddagger}$ & A1V & 25 Myr & \citet{Young04}& \citet{Jeffr05}\\
NGC2547/id8 & G7V & 35 Myr & \citet{Gorl07} & \citet{Jeffr05}\\ 
NGC2547/id7 & M & 35 Myr &\citet{Gorl07} & \citet{Jeffr05}\\
P1121 & F9V & 80 Myr & \citet{Gorl04} & \citet{Rojo97} \\
BD+20 307 & G0 & 2 Gyr & \citet{Song05} & \citet{Song05} \\
HD23514 & F6 & 120 Myr &\citet{Rhee08} & \citet{Rhee08}  \\
HD181296 &A0V & 12Myr &\citet{Su06} & \citet{Zuck04b}\\
HD62938 (\#12) & A0V &50-80 Myr & this work & \citet{Plat01}\\
\enddata
\tablenotetext{$\dagger$}{excess comes from free-free emission}
\tablenotetext{$\ddagger$}{incorrectly listed as a Cataclysmic Variable Star in SIMBAD}
\end{deluxetable*}

Fig \ref{fig:largeexc} shows the excess vs age diagram for our sample of 10 stars together with the sample of \citet{Riek05}. We express the excess  in 2.5$\times log(R)$ so we can compare the \citet{Riek05} data with other publications where the excess is expressed in K-[24] mag. All but three of the stars with large excesses fall within the envelope of the decay trend of \citet{Riek05}. (The normalization of this envelope is somewhat arbitrary so no distinction should be made between stars just above or just below it.) Two exceptions,  P1121 of \citet{Gorl04} and HD23514 of \citet{Rhee08}, are young. The third one BD+20 307 \citep{Song05} is very old (about 2 Gyr \citep{Zuck08}). The excesses of these stars are much higher than would be predicted based on their ages. Another two stars that are older than 30 Myr are borderline cases close to the upper envelope of the decay trend. One of them is a G7 star (ID8 in NGC 2547 \citet{Gorl07}) the other is an M star in the same cluster (ID7). We retain these two stars as candidate extreme excess stars. 

Although these stars were identified only on the basis of their large excesses at 24$\mu$m, they have a number of unique characteristics in comparison with other debris disks. For example, without full spectral energy distributions we cannot compute the fractional luminosities, L$_d$/L$_*$, but we can put lower limits on this parameter. As an illustration, we assume a star represented by a 6000K blackbody and a blackbody-like excess ten times the stellar output at 24 $\mu$m. The minimum fractional luminosity is then $2 \times 10^{-3}$ at an excess temperature of between 150 and 160K. Seven of the ten large excess stars are at this level or above. The minimum fractional luminosity for the remaining three large excess sources is $ \geq 2 \times 10^{-4}$. For comparison, debris disks typically have fractional luminosities below 10$^{-4}$ \citep{Moor06}. In addition, many of these systems are known to have strong spectral features in the mid-infrared, establishing that their emission there is dominated by warm, very finely divided dust ($\beta$ Pic, \citet{Chen07}; NGC 2547 ID8 and P1121, Gorlova et al. in preparation; BD +20 307, \citet{Song05}; HD 23514, \citet{Rhee08}). 

These latter sources represent extreme examples of debris disks containing warm dust. Additional examples of systems with warm dust include Fomalhaut \citep{Stape04}; $\beta$ Leo, \citep[][our unpublished observations]{Akes09}; $\eta$ Tel, \citep{Smith09}; $\eta$ Corvi, \citep{Smith08}; HR 4796A, \citep{Wahha05}; $\epsilon$ Eri, \citep{Backm09}; HD 113766, \citep{Lisse08}; HD 69830, \citep{Beich05}; and $\zeta$ Lep, \citep{Moerc07}. \citet{Moral09} report 28 additional systems with dust at T $\geq$ 180 K and that dominates the disk emission at 24 $\mu$. It is therefore plausible that many of the other large excess systems have substantial emission by warm dust.

Although many of the large-excess stars are roughly of solar type, even in this case they are relatively uncommon. In particular, for the 30 - 130 Myr age range,  Table \ref{tabl:largeexc} shows there are only three solar-like extreme excess stars. They are drawn from a parent population of all the stars observed within this age range and with spectral types between mid-F (F4) and mid-K (K4), a total of about 250 (from M47, \citet{Gorl04}; the Pleiades, \citet{Stau05}, \citet{Gorl06}, and our unpublished work; IC 2391, \citet{Sieg07}; NGC 2547, \citet{Gorl07}; the FEPS sample, \citet{Meye08}; and NGC 2451A\&B, this work).  That is, the incidence of extreme excess systems among solar-like stars in the 30-130Myr age range is only about 1\%. 

The 30-130Myr age range insures that these extreme excesses are unlikely to originate in the wave of oligarth planet building that peaks around 10-15 Myr (e.g., \citet{Curr08} and references therein). There are alternative possibilities for generating them, however. One is collisions between large bodies, similar to the collision that led to the formation of the Moon \citep{Rhee08}. It is thought that about three moon masses of material were thrown into orbit around the Sun in the impact that formed our Moon \citep{Canu04}. The rapid injection of such a large mass would produce a signature roughly in agreement with the observations of these large-excess objects (Gorlova et al. in preparation). A rough e-folding time of 2 million years for the dissipation of this mass can be estimated from the calculations of \citet{Grog01}.  In addition, a process that results in very vigorous gravitational stirring of a dense asteroid belt (e.g., through gravitationally resonant orbits of massive planets, \citet{Gome05}) is also a candidate to produce huge infrared excesses.   

Although Table \ref{tabl:largeexc} contains several early type stars older than 20 Myr and with large excesses, none of them is above the decay trend of normal debris disk stars. In fact, as Table \ref{tabl:largeexc} emphasizes, all the large-excess early type stars are relatively young and may just still be in the decay of their initial peak debris output. The placement of the trend line is rather arbitrary. Nevetheless, we have considered why no very large excesses are found for older early-type stars. We calculated what mass would be needed to cause an excess about 4 times above the stellar photosphere at 24 $\mu$m in an A star. We used only grains larger than 6.3 $\mu$m, because radiation pressure will eject small grains from the system, with terminal velocity $v = {2 G M_{\star} \over r_{init}} (\beta - {1 \over 2})^{1 \over 2}$ where $M_{\star}$ is the mass of the central star, $r_{init}$ is the radius of the initial orbit of the grain and $\beta$ is the ratio of forces from radiation pressure and gravity. \citet{Domi03} showed that the orbiting grains normally dominate the visibility of the debris disk over the blown-out grains. To get the total mass needed to cause the required 24$\mu$m excess we first assumed that all the flux comes from the smallest grains then we calculated the total surface area of these grains. Using the surface area and the $N(a) \sim a^{-3.5}$ grain size distribution we were able to determine the scaling constant using the following equation

\begin{equation}
\int^{a_{max}}_{a_{min}} 4 \pi {{a^2} \over 4} c a^{-3.5} da = N_{a_{min}} 4 \pi {{a^2_{min}} \over 4}
\end{equation}

\noindent where $a_{min}$ is the diameter of the smallest grain remaning in the system and $a_{max}$ is a nominal maximum parent planetesimal size (100m). Using the scaling constant we can calculate the total mass by integrating ${4 \over 3} \pi  a^3 \rho c a^{-3.5}$ from $a_{min}$ to $a_{max}$. We found that the total mass needed to produce the excess is on the order of a couple of Moon masses. The same calculation for G-type stars with $a_{min} = 0.01 \mu$m leads to two orders of magnitude less mass ($\sim10^{-2} M_{Moon}$).

In the case of A stars, radiation pressure clears the small grains from the disk in a few hundred years while these grains remain around the G-star for a few times $10^4$ years \citep{Chen05}. To quantify the effects of this difference, we also calculated the scaling factor and the total mass neglecting the effect of blowout for the A star and integrating over the entire size range from $10^{-2} \mu$m to 100m. We found that in this case we would need more than 25 times less material for producing the same level of excess around the A star. In this hypothetical case the masses needed for A- and G-type stars are the same order of magnitude. This implies that  their ability to eject grains very efficiently through photon pressure plays a large role in the absence of extreme excess around older early type stars.Those small grains carry a large surface area (relative to their mass) which would make the excess more easily detectable. A corollary is that large grain production rates around early-type stars will be indicated by large flows of small grains out of the system driven by photon pressure. Possible examples include Vega \citep{Su05} and HR8799 (Su et al. in preparation)

\section{Conclusions}
We present optical and {\it Spitzer}/IRAC-MIPS photometry and medium resolution spectroscopy of the central 0.75 sq deg. region of NGC~2451 A and B. Using radial velocity data and optical/near-IR color magnitude diagrams we selected 61 members of NGC~2451 A and 131 members of NGC2451 B. We found one object in NGC2451 B and none in NGC2451 A with excess at 8$\mu$m. We conclude that 8 micron excesses at the ages of these clusters are probably associated with an isolated event that disturbs the distribution of planetesimals to yield a sequence of collisional cascades producing dust.

We find one object with 24$\mu$m excess among the early type stars of NGC 2451 B and none in NGC2451 A. For the solar type stars we found 5 out of 15 stars with excess in NGC2451 A and 7 out of 16 for NGC2451 B. In general these results agree with excess fractions reported in the literature. 

We used the data presented in \citet{Gasp08a} supplemented with our new datapoints to plot the excess fraction vs age, and found that the excess fraction unexpectedly drops in the range of 30-80Myr for early type stars while it slightly rises in the same age for late type stars. We tested our findings by binning the cluster data of \citet{Gasp08a} into equal logarithmic age bins. The results are consistent with some structure but also indicate that the fluctuations may be statistical in nature. 

We summarize the detections to date of large debris-disk 24 $\mu$m excesses. There are only five extreme cases, with fractional luminosities $\geq$ $2 \times 10^{-3}$. We did not find extreme excesses around older early type stars, excesses that would indicate a catastrophic collison between forming planetesimals. We found that blowout of the small grains probably plays a very important role in the lack of detections. 

\acknowledgments

We thank the anonymous referee for comments and suggestions that greatly improved the manuscript. This work is based on observations made with the Spitzer Space Telescope, which is operated by the Jet Propulsion Laboratory, California Institute of Technology, under NASA contract 1407. Support for this work was provided by NASA through contract 1255094, issued by JPL/Caltech. ZB also received support from Hungarian OTKA Grants T049082 and K76816. LLK has been supported by the Anglo-Australian Observatory and the Australian Research Council

\clearpage

\clearpage
\LongTables
\begin{landscape}
\setcounter{table}{2}
\begin{deluxetable}{llcccccccccccccccc}
\tablecolumns{18}
\tabletypesize{\scriptsize} 
\tablewidth{0pc}
\tablecaption{UBVRIJHK Photometry of the members of NGC2451 A.\label{tab:membAubvri}}
\tablehead{
\colhead{RA}&\colhead{DEC}&\colhead{U}&\colhead{$\sigma$(U)}&\colhead{B}&\colhead{$\sigma$(B)}&\colhead{V}&\colhead{$\sigma$(V)}&\colhead{R}&\colhead{$\sigma$(R)}&\colhead{I}&\colhead{$\sigma$(I)}&\colhead{$\sigma$(J)}&\colhead{$\sigma$(J)}&\colhead{H}&\colhead{$\sigma$(H)}&\colhead{K}&\colhead{$\sigma$(K)}\\
\colhead{J2000}&\colhead{J2000}&\colhead{mag}&\colhead{mag}&\colhead{mag}&\colhead{mag}&\colhead{mag}&\colhead{mag}&\colhead{mag}&\colhead{mag}&\colhead{mag}&\colhead{mag}&\colhead{mag}&\colhead{mag}&\colhead{mag}&\colhead{mag}&\colhead{mag}&\colhead{mag}
}
\startdata
115.72653 & -37.99071 & \nodata&\nodata & \nodata&\nodata & 19.841 & 0.043 & 18.399&0.017&16.579&0.014 & 14.793&0.031  & 14.201&0.038  & 13.917&0.059 \\ 
115.73360 & -38.35236 & \nodata&\nodata & \nodata&\nodata & 19.495 & 0.035 & 17.700&0.016&15.724&0.012 & 13.626&0.032  & 13.061&0.034  & 12.712&\nodata \\ 
115.75735 & -37.70676 & \nodata&\nodata & \nodata&\nodata & 19.095 & 0.036 & 17.701&0.017&16.180&0.009 & 14.628&0.032  & 14.010&0.033  & 13.685&0.060 \\ 
115.77562 & -37.69178 & \nodata&\nodata & \nodata&\nodata & 18.338 & 0.017 & 16.937&0.020&15.265&0.009 & 13.666&0.031  & 13.076&0.028  & 12.782&0.032 \\ 
115.77648 & -38.15585 & 17.636&0.013 & 16.611&0.006 & 15.152 & 0.008 & 14.158&0.010&13.017&0.008 & 11.758&0.023  & 11.104&0.020  & 10.912&0.021 \\ 
115.80866 & -38.21898 & \nodata&\nodata & \nodata&\nodata & 18.941 & 0.023 & 17.641&0.008&16.050&0.006 & 14.443&0.037  & 13.792&0.044  & 13.537&0.043 \\ 
115.82023 & -38.04878 & \nodata&\nodata & \nodata&\nodata & 19.826 & 0.046 & 18.482&0.013&16.752&0.009 & 15.153&0.043  & 14.527&0.059  & 14.209&0.069 \\ 
115.84371 & -37.73632 & \nodata&\nodata & 19.111&0.034 & 17.840 & 0.013 & 16.603&0.007&15.135&0.006 & 13.700&0.026  & 13.046&0.020  & 12.732&0.028 \\
\enddata
\tablenotetext{~}{numbers next to the coordinates designate 24 $\mu$m excees sources from Table \ref{tab:24exces}}
\end{deluxetable}

\begin{deluxetable}{llcccccccccccccccc}
\tablecolumns{18}
\tabletypesize{\scriptsize} 
\tablewidth{0pc}
\tablecaption{UBVRIJHK Photometry of the members of NGC2451 B\label{tab:membBubvri}}
\tablehead{
\colhead{RA}&\colhead{DEC}&\colhead{U}&\colhead{$\sigma$(U)}&\colhead{B}&\colhead{$\sigma$(B)}&\colhead{V}&\colhead{$\sigma$(V)}&\colhead{R}&\colhead{$\sigma$(R)}&\colhead{I}&\colhead{$\sigma$(I)}&\colhead{$\sigma$(J)}&\colhead{$\sigma$(J)}&\colhead{H}&\colhead{$\sigma$(H)}&\colhead{K}&\colhead{$\sigma$(K)}\\
\colhead{J2000}&\colhead{J2000}&\colhead{mag}&\colhead{mag}&\colhead{mag}&\colhead{mag}&\colhead{mag}&\colhead{mag}&\colhead{mag}&\colhead{mag}&\colhead{mag}&\colhead{mag}&\colhead{mag}&\colhead{mag}&\colhead{mag}&\colhead{mag}&\colhead{mag}&\colhead{mag}
}
\startdata
115.72778 & -37.95026 & 16.698 & 0.014 & 15.763 & 0.005 & 14.607 & 0.010 & 13.908 & 0.020 & 13.281 & 0.025 & 12.465 & 0.018  & 11.919 & 0.020  & 11.773 & 0.019 \\ 
115.75046 & -37.79300 & 19.520 & 0.072 & 18.393 & 0.020 & 16.887 & 0.007 & 15.726 & 0.006 & 14.473 & 0.004 & 13.047 & 0.026  & 12.345 & 0.026  & 12.106 & 0.024 \\ 
115.75243 & -37.95836 & 13.070 & 0.009 & 12.958 & 0.004 & 12.332 & 0.008 & 11.940 & 0.013 & 11.542 & 0.019 & 11.140 & 0.025  & 10.794 & 0.020  & 10.757 & 0.023 \\ 
115.77086 & -37.81706 & 12.230 & 0.004 & 12.070 & 0.002 & 11.563 & 0.002 & 11.199 & 0.002 & 10.918 & 0.004 & 10.425 & 0.023  & 10.217 & 0.020  & 10.134 & 0.019 \\ 
115.77200 & -37.98902 & 18.918 & 0.042 & 17.743 & 0.009 & 16.189 & 0.007 & 15.040 & 0.009 & 13.623 & 0.012 & 12.358 & 0.025  & 11.774 & 0.020  & 11.533 & 0.023 \\ 
115.79673 & -38.09564 & \nodata & \nodata & 15.036 & 0.003 & 14.024 & 0.004 & 13.427 & 0.008 & 12.869 & 0.007 & 12.031 & 0.021  & 11.503 & 0.020  & 11.406 & 0.019 \\ 
115.79724 & -38.15298 & 17.356 & 0.012 & 16.408 & 0.005 & 15.099 & 0.007 & 14.359 & 0.008 & 13.644 & 0.008 & 12.551 & 0.023  & 11.965 & 0.020  & 11.772 & 0.017 \\ 
115.80643 & -37.79030 & \nodata & \nodata & 20.087 & 0.084 & 18.473 & 0.020 & 17.077 & 0.007 & 15.387 & 0.007 & 13.659 & 0.026  & 13.062 & 0.028  & 12.764 & 0.032 \\
\enddata
\tablenotetext{~}{numbers next to the coordinates designate 24 $\mu$m excees sources from Table \ref{tab:24exces}}
\end{deluxetable}

\begin{deluxetable}{llcccccccccc}
\tablecolumns{12}
\tabletypesize{\scriptsize} 
\tablewidth{0pc}
\tablecaption{{\it Spitzer} 3.6$\mu$m - 24$\mu$m photometry of mebers of NGC2451 A\label{tab:membAspitz}}
\tablehead{
\colhead{RA}&\colhead{DEC}&\colhead{[3.6]}&\colhead{$\sigma$([3.6])}&\colhead{[4.5]}&\colhead{$\sigma$([4.5])}&\colhead{[5.8}&\colhead{$\sigma$([5.8])}&\colhead{[8.0]}&\colhead{$\sigma$([8.0])}&\colhead{[24]}&\colhead{$\sigma$([24])}\\
\colhead{J2000}&\colhead{J2000}&\colhead{mag}&\colhead{mag}&\colhead{mag}&\colhead{mag}&\colhead{mag}&\colhead{mag}&\colhead{mag}&\colhead{mag}&\colhead{mag}&\colhead{mag}
}
\startdata
115.72653 & -37.99071 & \nodata & \nodata & \nodata & \nodata & \nodata & \nodata & \nodata & \nodata & \nodata & \nodata \\ 
115.73360 & -38.35236 & \nodata & \nodata & \nodata & \nodata & \nodata & \nodata & \nodata & \nodata & \nodata & \nodata \\ 
115.75735 & -37.70676 & \nodata & \nodata & \nodata & \nodata & \nodata & \nodata & \nodata & \nodata & \nodata & \nodata \\ 
115.77562 & -37.69178 & \nodata & \nodata & \nodata & \nodata & \nodata & \nodata & \nodata & \nodata & \nodata & \nodata \\ 
115.77648 & -38.15585 & \nodata & \nodata & \nodata & \nodata & \nodata & \nodata & \nodata & \nodata & \nodata & \nodata \\ 
115.80866 & -38.21898 & \nodata & \nodata & \nodata & \nodata & \nodata & \nodata & \nodata & \nodata & \nodata & \nodata \\ 
115.82023 & -38.04878 & \nodata & \nodata & \nodata & \nodata & \nodata & \nodata & \nodata & \nodata & \nodata & \nodata \\ 
115.84371 & -37.73632 & \nodata & \nodata & \nodata & \nodata & \nodata & \nodata & \nodata & \nodata & \nodata & \nodata \\
\enddata
\tablenotetext{~}{numbers next to the coordinates designate 24 $\mu$m excees sources from Table \ref{tab:24exces}}
\end{deluxetable}

\begin{deluxetable}{llcccccccccc}
\tablecolumns{12}
\tabletypesize{\scriptsize} 
\tablewidth{0pc}
\tablecaption{{\it Spitzer} 3.6$\mu$m - 24$\mu$m photometry of mebers of NGC2451 B\label{tab:membBspitz}}
\tablehead{
\colhead{RA}&\colhead{DEC}&\colhead{[3.6]}&\colhead{$\sigma$([3.6])}&\colhead{[4.5]}&\colhead{$\sigma$([4.5])}&\colhead{[5.8}&\colhead{$\sigma$([5.8])}&\colhead{[8.0]}&\colhead{$\sigma$([8.0])}&\colhead{[24]}&\colhead{$\sigma$([24])}\\
\colhead{J2000}&\colhead{J2000}&\colhead{mag}&\colhead{mag}&\colhead{mag}&\colhead{mag}&\colhead{mag}&\colhead{mag}&\colhead{mag}&\colhead{mag}&\colhead{mag}&\colhead{mag}
}
\startdata
115.72778 & -37.95026 & \nodata & \nodata & \nodata & \nodata & \nodata & \nodata & \nodata & \nodata & \nodata & \nodata \\ 
115.75046 & -37.79300 & \nodata & \nodata & \nodata & \nodata & \nodata & \nodata & \nodata & \nodata & \nodata & \nodata \\ 
115.75243 & -37.95836 & \nodata & \nodata & \nodata & \nodata & \nodata & \nodata & \nodata & \nodata & \nodata & \nodata \\ 
115.77086 & -37.81706 & \nodata & \nodata & \nodata & \nodata & \nodata & \nodata & \nodata & \nodata & \nodata & \nodata \\ 
115.77200 & -37.98902 & \nodata & \nodata & \nodata & \nodata & \nodata & \nodata & \nodata & \nodata & \nodata & \nodata \\ 
115.79673 & -38.09564 & \nodata & \nodata & \nodata & \nodata & \nodata & \nodata & \nodata & \nodata & \nodata & \nodata \\ 
115.79724 & -38.15298 & \nodata & \nodata & \nodata & \nodata & \nodata & \nodata & \nodata & \nodata & \nodata & \nodata \\ 
115.80643 & -37.79030 & \nodata & \nodata & \nodata & \nodata & \nodata & \nodata & \nodata & \nodata & \nodata & \nodata \\
\enddata
\tablenotetext{~}{numbers next to the coordinates designate 24 $\mu$m excees sources from Table \ref{tab:24exces}}
\end{deluxetable}
\clearpage
\end{landscape}

\end{document}